\documentclass[useAMS]{mn2e}
\usepackage{afterpage}
\usepackage{longtable}
\usepackage{times}
\usepackage[]{natbib}
\usepackage[dvips]{graphicx}
\usepackage{amssymb,amsmath}

\def\apj{ApJ}                 
\def\apjl{ApJ}                
\def\aap{A\&A}                
\def\mnras{MNRAS}             
\def\nat{Nature}              

\font \bolditalics = cmmib10
\def \vc #1{{\textfont1=\bolditalics \hbox{$\bf#1$}}}

\def\Cg{{\bf C}}
\def\mg{{\bf m}}
\def\Fg{{\bf F}}
\def\F{F}

\def\Msun{M_{\odot}}

\def\NiT{N_{\rm i}^{\rm T}}

\def\NkT{N_{\rm k}^{\rm T}}
\def\Nko{N_{\rm k}^{\rm o}}
\def\NlT{N_{\rm l}^{\rm T}}
\def\Nio{N_{\rm i}^{\rm o}}
\def\Njo{N_{\rm j}^{\rm o}}
\def\fii{f_{\rm ii}}
\def\fjj{f_{\rm jj}}
\def\fkk{f_{\rm kk}}
\def\fij{f_{\rm ij}}
\def\fji{f_{\rm ji}}
\def\fkj{f_{\rm kj}}
\def\fki{f_{\rm ki}}
\def\fik{f_{\rm ik}}
\def\fjk{f_{\rm jk}}
\def\flj{f_{\rm lj}}

\def\wijo{\omega_{\rm ij}^{\rm o}}
\def\wjio{\omega_{\rm ji}^{\rm o}}
\def\wij{\omega_{\rm ij}}

\def\wiio{\omega_{\rm ii}^{\rm o}}
\def\wjjo{\omega_{\rm jj}^{\rm o}}
\def\wiiT{\omega_{\rm ii}^{\rm T}}

\def\wkkT{\omega_{\rm kk}^{\rm T}}
\def\wkko{\omega_{\rm kk}^{\rm o}}
\def\wklT{\omega_{\rm kl}^{\rm T}}

\def\Dijo{(D_{\rm{i}}D_{\rm{j}})_\theta^{\rm{o}}}
\def\DklT{(D_{\rm{k}}D_{\rm{l}})_\theta^{\rm{T}}}
\def\Dij{(D_{\rm{i}}D_{\rm{j}})_\theta}

\def\RR{(RR)_\theta}
\def\DiR{(D_{\rm{i}}R)_\theta}
\def\DjR{(D_{\rm{j}}R)_\theta}
\def\Nr{N_{\rm R}}
\def\Ni{N_{\rm i}}
\def\Nj{N_{\rm j}}

\def\foo{f_{\rm 11}}
\def\fww{f_{\rm 22}}

\def\fow{f_{\rm 12}}

\def\fwo{f_{\rm 21}}

\def\wowo{\omega_{\rm 12}^{\rm o}}
\def\wooo{\omega_{\rm 11}^{\rm o}}
\def\wwwo{\omega_{\rm 22}^{\rm o}}

\def\Noo{N_{\rm 1}^{\rm o}}
\def\Nwo{N_{\rm 2}^{\rm o}}
\def\NoT{N_{\rm 1}^{\rm T}}
\def\NwT{N_{\rm 2}^{\rm T}}

\begin{document}
\title[Photometric redshift contamination]{Photometric redshifts: estimating their contamination and distribution using clustering information}
\author[Benjamin et al.]{Jonathan Benjamin$^{1}$\thanks{jonben@phas.ubc.ca}, Ludovic Van Waerbeke$^{1}$, Brice M\'enard$^{2}$, Martin Kilbinger$^{3,4,5}$\\
	$^1$ University of British Columbia, 6224 Agricultural Road, Vancouver, V6T 1Z1, B.C., Canada.\\
	$^2$ Canadian Institute for Theoretical Astrophysics, University of Toronto, 60 George Street, Toronto, ON, M5S 3H4, Canada.\\
	$^3$ Excellence Cluster Universe, Technische Universit\"at M\"unchen, Boltzmannstr. 2, 85748 Garching, Germany.\\
	$^4$ Universit\"ats-Sternwarte M\"unchen, Scheinerstr. 1, D-81679 M\"unchen, Germany.\\
	$^5$  Institut d'Astrophysique de Paris, UMR 7095 CNRS \&
 Universit\'e Pierre et Marie Curie, 98 bis boulevard Arago, 75014 Paris, France.}

\maketitle

\begin{abstract}
We present a new technique to estimate the level of contamination between photometric redshift bins. If the true angular cross-correlation between redshift bins can be safely assumed to be zero, any measured cross-correlation is a result of contamination between the bins. We present the theory for an arbitrary number of redshift bins, and discuss in detail the case of two and three bins which can be easily solved analytically. We use mock catalogues constructed from the Millennium Simulation to test the method, showing that artificial contamination can be successfully recovered with our method. We find that degeneracies in the parameter space prohibit us from determining a unique solution for the contamination, though constraints are made which can be improved with larger data sets. We then apply the method to an observational galaxy survey; the deep component of the Canada France Hawaii Telescope Legacy Survey. We estimate the level of contamination between photometric redshift bins and demonstrate our ability to reconstruct both the true redshift distribution and the true average redshift of galaxies in each photometric bin.
\end{abstract}

\begin{keywords}
 galaxies: distances and redshifts - galaxies: photometry - techniques: photometric - methods: analytical - large-scale structure of Universe
\end{keywords}

\section{Introduction}
The time-intensive nature of spectroscopy and the immense number of galaxies in current and future surveys have secured the place of photometric redshifts among the most valuable astronomical tools. Nearly all cosmological probes are sensitive to distance, making redshifts particularly important to cosmology. While spectroscopy is the most accurate and precise way to measure redshift, it is far too time-intensive to apply to current or future surveys which require redshift estimates for millions of galaxies. Hence, photometric redshifts and a thorough understanding of their uncertainties are vital to the study of large galaxy surveys.

The observed colours of a galaxy are often unable to uniquely specify a galaxy type and redshift. Degeneracies in fitting galaxy colours can result in significant misidentifications causing contamination between all redshifts.   In this paper we investigate contamination between photometric redshift bins and present a method for estimating the contamination using the angular correlation function. This method relies only on the photometric redshifts, it does not require a spectroscopic sample. However, it is likely that the photometric redshifts will have been calibrated with a spectroscopic training sample.

Quantifying photometric redshift contamination is essential to exploit the full potential of future photometric surveys, such as the Large Synoptic Survey Telescope (LSST) or the Supernova Acceleration Probe (SNAP), which aim to constrain dark energy. Weak lensing and baryon acoustic oscillations (BAO) are both sensitive to the mean redshift of a bin, requiring an unbiased measurement on the order of $0.001-0.005$ in redshift so that the constraints on dark energy are not significantly degraded \citep{Huterer2004}.

Several studies have investigated the use of the angular correlation function in determining the true redshift distribution of galaxies binned by photometric redshift \citep{Schneider2006,Newman2008,Zhang2009}. These works have adopted LSST-like survey parameters and focused on the Fisher information matrix as a means of forecasting constraints. \citet{Schneider2006} show that the angular correlation function in the linear regime can be used to measure the mean redshift to an accuracy of 0.01, noting that there is further constraining power at smaller scales. The method investigated by \citet{Newman2008} relies on an overlapping spectroscopic sample so that the angular cross-correlation between it and the photometric redshift bins can be exploited. They find that the desired accuracy on the mean redshift can be reached provided a spectroscopic sample of 25,000 galaxies per unit redshift. \citet{Zhang2009} investigate a self-calibration method, where the mean redshift in a bin is estimated from angular cross-correlations between photometric redshift bins, very similar to the one presented in this work and \citet{Erben2009}. They demonstrate that self-calibration can reach the required accuracy on the mean redshift if additional information from weak lensing shear is used to help break parameter degeneracies.

We focus here on the practical application of measuring photometric contamination in both simulated and real data. This is of interest not only for meeting the stringent requirements of future surveys as mentioned above, but also for other applications such as: a diagnostic tool for photometric redshifts; determining background samples for cluster lensing and estimating the true redshift distribution for 2-D cosmic shear measurements. The details of our method for a strict two-bin analysis of the CFHTLS-Archive-Research-Survey (CARS) are presented in \citet{Erben2009}, where we demonstrate that the contamination present in the bright spectroscopic training sample is consistent with the contamination seen in the much deeper photometric redshift sample. This addresses a central concern when calibrating photometric redshifts with spectroscopic redshifts.

This paper is organised as follows: The angular correlation function and the estimators we use are presented in \S\ref{sec:angcorr}. The details of the analytic method for estimating redshift bin contamination are discussed in \S\ref{sec:theory}, which first addresses the general problem of contamination between an arbitrary number of redshift bins before focusing on the two-bin case and its extension to multiple redshift bins. The analytic model is tested using mock observational catalogues in \S\ref{sec:MilSim}, where we show that contamination between redshift bins can be accurately determined with our model. In \S\ref{sec:Deep} we measure contamination in a real galaxy survey demonstrating the ability of the method to constrain the true (uncontaminated) redshift distribution and measure the average redshift of each photometric redshift bin. Concluding remarks, including limitations of the method and ideas for future work, are presented in \S\ref{sec:discussion}. Details of the maximum likelihood method, including a detailed discussion of the covariance matrix, is left to appendix \ref{sec:covariance}. The contamination model for three redshift bins is discussed in appendix \ref{sec:threebin}.

\section{Angular Correlation Function and Estimators}
\label{sec:angcorr}
The two-point angular correlation function describes the amount of clustering in a distribution of galaxies relative to what would be expected from a random distribution. The angular correlation function $\omega$ can be interpreted as the excess probability of finding an object in the solid angle ${\rm d}\Omega$ a distance $\theta$ from another object. The total probability is given by
\begin{equation}
 dP=N[1+\omega(\theta)]{\rm d}\Omega,
\end{equation}
\noindent where $N$ is the density of objects per unit solid angle \citep{Peebles1980}. Note that $N {\rm d}\Omega$ is the Poisson probability that the solid angle element is occupied by a galaxy.

Many estimators have been devised for the correlation function. \citet{KSS2000} present a comparison of the most widely used estimators. They mainly differ in the handling of edge effects, and for arbitrarily large number densities they all converge. In \S\ref{sec:theory} we employ the simplest estimator,
\begin{equation}
\wij=\frac{\Dij}{\RR}\frac{\Nr\Nr}{\Ni\Nj} - 1, \label{eq:angcorr_norm}
\end{equation}
\noindent where $\Dij$ is the number of pairs separated by a distance $\theta$ between data sets i and j, $\RR$ is the number of pairs separated by a distance $\theta$ for a random set of points, $\Nr$ is the number of points in the random sample and $\Ni$ ($\Nj$) is the number of points in data sample i (j). In this work we consider i and j to be non-overlapping redshift bins. The auto-correlation is described by the case i$=$j, and the cross-correlation by the case i$\neq$j.

A more robust estimator \citep{L&S1993} is
\begin{equation}
\wij=\frac{\Dij}{\RR}\frac{\Nr\Nr}{\Ni\Nj} - \frac{\DiR}{\RR}\frac{\Nr}{\Ni}-\frac{\DjR}{\RR}\frac{\Nr}{\Nj} + 1. \label{eq:angcorr_LS}
\end{equation}
\noindent This is used when measuring the correlation function from data, either the Millennium Simulation in \S\ref{sec:MilSim} or the CFHTLS-Deep fields in \S\ref{sec:Deep}.

Galaxies cluster in over-dense regions, leading to an excess number of pairs when compared to a random distribution of points. On small scales, the angular auto-correlation function of galaxies is positive for all redshifts, though the shape and amplitude vary as a function of redshift due to the evolution of structure formation. The angular cross-correlation between two distant redshift bins is zero since galaxies that are physically separated by large distances do not cluster with one another.

When considering galaxies binned in redshift, neighbouring bins may have a significant cross-correlation, especially if the bin width is narrow, since a significant number of galaxies in each bin could be clustered with each other. In the case of photometric redshifts the typical redshift error ($\Delta z \approx 0.05$) will result in a large cross-correlation if the width of the bins is not much larger than this error. 

Photometric redshift bins that are not neighbouring should not be physically clustered with one another and their cross-correlation should be zero. Deviations from zero indicate that the two bins contain galaxies that are physically clustered. These galaxies may result from contamination between the two bins or from the mutual contamination of both bins from other redshifts.

Weak lensing magnification can cause galaxies at high redshift to cluster near lower redshift galaxies. This effect can be calculated and accounted for --we discuss this in \S\ref{sec:discussion}.

\section{Analytic Development of the Contamination Model}
\label{sec:theory}
The goal of the contamination model is to measure the level of contamination between all photometric redshift bins by measuring the angular cross-correlation. We have presented the details of the two-bin model in \citet{Erben2009} and applied it to the CFHTLS-Archive-Research Survey (CARS). Here we develop a fully consistent multi-bin approach in \S\ref{sec:MultiBin} before revisiting the two-bin case and extending it to a global pairwise analysis in \S\ref{sec:TwoBins}.

\subsection{Multi-Bin Analysis}
\label{sec:MultiBin}
We first address the case of an arbitrary number of redshift bins, where each bin is potentially contaminating every other bin. The number of galaxies observed to be in the ${\rm i^{\rm th}}$ bin is $\Nio$ which, by virtue of the mixing between bins, is not equal to the true number of galaxies in that redshift bin $\NiT$. We define $\fij$ to be the number of galaxies from bin i contaminating bin j as a fraction of the true number of galaxies in bin i. Therefore $\fij \NiT$ is the number of galaxies from bin i present in bin j. The fraction of galaxies in bin i which do not contaminate other bins is taken to be $\fii$ which is convenient shorthand for
\begin{equation}
 \fii=1-\sum^{\rm n}_{\rm k\neq i}\fik, \label{eq:fii}
\end{equation}
\noindent where n is the number of redshift bins.

The observed number of galaxies in bin i will be those galaxies which contaminate the bin, plus those galaxies from bin i which do not contaminate other bins,
\begin{equation}
 \Nio=\sum^{\rm n}_{\rm k}\NkT\fki, \label{eq:nbinNio}
\end{equation}
\noindent where the k=i term accounts for those galaxies that remain in bin i while all other terms account for galaxies that have contaminated bin i. This results in a system of n equations which can be written as:
\begin{equation}
\begin{pmatrix}
N_{\rm 1}^{\rm o} \\
N_{\rm 2}^{\rm o} \\
\hdotsfor{1} \\
N_{\rm n}^{\rm o}
\end{pmatrix}=
\begin{pmatrix}
f_{\rm 11} & f_{\rm 21} & \dots & f_{\rm n1} \\
f_{\rm 12} & f_{\rm 22} & \dots & f_{\rm n2} \\
\hdotsfor{4} \\
f_{\rm 1n} & f_{\rm 2n} & \dots & f_{\rm n n} \\
\end{pmatrix}
\begin{pmatrix}
N_{\rm 1}^{\rm T} \\
N_{\rm 2}^{\rm T} \\
\hdotsfor{1} \\
N_{\rm n}^{\rm T}
\end{pmatrix}.\label{eq:Niomatrix}
\end{equation}

If the $\rm n\times n$ matrix is invertible then we can solve the system for the true number of galaxies in each bin. In the limit of zero contamination the off-diagonal elements will tend to zero while the diagonal elements tend to unity. In this limit the matrix is trivially non-singular. If a matrix is strictly diagonally dominant then it follows from the Gershgorin circle theorem that it is non-singular. Therefore as long as 
\begin{equation}
\fii > \sum^{\rm n}_{\rm k \neq i}\fik,
\end{equation}
\noindent the matrix is strictly diagonally dominant and therefore is invertible. This condition is simply stating that a solution exits if the majority of the galaxies from the $\rm i^{th}$ true redshift bin do not contaminate other bins. The case of uniform contamination, where each bin sends $\NiT/{\rm n}$ galaxies to each other bin, results in a singular matrix where all rows are identical.

The observed correlation functions can be derived by investigating which pairs contribute when correlating bin i and bin j. The observed number of data pairs between bins can be related to contributions from the true number of pairs,
\begin{equation}
 \Dijo=\sum^{\rm n}_{\rm k}\sum^{\rm n}_{\rm l}\DklT\fki\flj. \label{eq:numpairs}
\end{equation}
\noindent Multiplying both sides by $\frac{1}{\RR}\frac{\Nr\Nr}{\Ni\Nj}$ and using Eq.~(\ref{eq:angcorr_norm}) to relate the pair counts to the correlation function yields
\begin{equation}
\wijo=\sum^{\rm n}_{\rm k}\sum^{\rm n}_{\rm l}\wklT\frac{\NkT\NlT}{\Nio\Njo}\fki\flj, \label{eq:nbin_angcorr}
\end{equation}
\noindent which can also be derived by considering $\left<\Nio,\Njo\right>$. 

We assume that the true cross-correlation between any two redshift bins is zero. Hence, it is useful to rewrite Eq.~ (\ref{eq:nbin_angcorr}) as
\begin{equation}
\wijo=\sum^{\rm n}_{\rm k}\wkkT\frac{(\NkT)^2}{\Nio\Njo}\fki\fkj.\label{eq:nbin_wijo}
\end{equation}
\noindent Note that this allows us to express the observed auto- and cross-correlations as linear combinations of the true auto-correlation functions. The following matrix equation follows from the above when we consider only the equations for the observed auto-correlations (i.e, when i=j):
\begin{equation}
\begin{pmatrix}
\omega_{\rm 11}^{\rm o}N_{\rm 1}^{\rm o^2} \\
\omega_{\rm 22}^{\rm o}N_{\rm 2}^{\rm o^2} \\
\hdotsfor{1} \\
\omega_{\rm nn}^{\rm o}N_{\rm n}^{\rm o^2}
\end{pmatrix}=
\begin{pmatrix}
f_{\rm 11}^2 & f_{\rm 21}^2 & \dots & f_{\rm n1}^2 \\
f_{\rm 12}^2 & f_{\rm 22}^2 & \dots & f_{\rm n2}^2 \\
\hdotsfor{4} \\
f_{\rm 1n}^2 & f_{\rm 2n}^2 & \dots & f_{\rm n n}^2 \\
\end{pmatrix}
\begin{pmatrix}
\omega_{\rm 11}^{\rm T}N_{\rm 1}^{\rm T^2} \\
\omega_{\rm 22}^{\rm T}N_{\rm 2}^{\rm T^2} \\
\hdotsfor{1} \\
\omega_{\rm nn}^{\rm T}N_{\rm n}^{\rm T^2}
\end{pmatrix}.\label{eq:wiiomatrix}
\end{equation}
We are again left with an $\rm n\times n$ matrix which, when inverted, lets us express the true auto-correlations in terms of the observed auto-correlations. Once they are known, we can use Eq.~(\ref{eq:nbin_wijo}) to express the observed cross-correlations as a linear combination of observed auto-correlation functions,
\begin{equation}
\wijo=\sum^{\rm n}_{\rm k}\wkko\frac{(\Nko)^2}{\Nio\Njo}g_{\rm k}(f),\label{eq:cross}
\end{equation}
\noindent where $\rm i \neq j$ and $g_{\rm k}(f)$ is a complicated function of the contamination fractions. Note that the true number of objects from Eq.~(\ref{eq:nbin_wijo}) has cancelled with that from Eq.~(\ref{eq:wiiomatrix}). 

The $\rm n\times n$ matrix in Eq.~(\ref{eq:wiiomatrix}) above is very similar to that found in Eq.~(\ref{eq:Niomatrix}). The diagonal elements tend to unity as the contamination between bins tends to zero, resulting in a nearly diagonal matrix. The matrix will be strictly diagonally dominant and hence non-singular under the same condition found above, and for uniform contamination the matrix is singular. 

The total number of unknown parameters ($\fij$, where $\rm i \neq j$) is $\rm n(n - 1)$. This is clear since, in general, each of the n bins can contaminate each of the other bins ($\rm n-1$). Our goal is to constrain these parameters with Eq.~(\ref{eq:cross}), which yields only $\rm \frac{n(n-1)}{2}$ equations since $\wijo = \wjio$. If we could only measure the angular correlation function at one scale then it would be impossible to constrain the parameters. However, if we have two independent measurements of the cross-correlations at different scales, we double the number of equations and are able to constrain the problem. In practice the measurements of the angular correlation function at different scales are not independent, and we should strive to include as large a range of scales as possible. 

Eq.~(\ref{eq:cross}) relates the amplitude of the cross-correlation to a weighted sum of the auto-correlations. If the shape of all the auto-correlations were the same, then it would be impossible to distinguish their contributions to the cross-correlation, which would also have the same shape. For example if the auto-correlations were well described by power laws with the same slope. In such cases this method would yield completely degenerate parameter constraints.

\subsection{Pairwise analysis}
\label{sec:TwoBins}
Here we first consider the case of exactly two redshift bins. This provides a simple case with which to test our method using the Millennium Simulation in \S\ref{sec:MilSim}. We then expand this to multiple redshift bins by considering each pair in turn. We show that this global pairwise analysis will yield accurate estimates for the entire set of contamination fractions if the contamination is sufficiently small.

We consider two redshift bins, labelled 1 and 2. From Eq.~(\ref{eq:Niomatrix}), we can express the observed number of galaxies in each bin in terms of the true number of galaxies and the contamination fractions,
\begin{eqnarray}
 \Noo&=&\NoT(1-\fow) + \NwT\fwo,\nonumber\\
 \Nwo&=&\NwT(1-\fwo) + \NoT\fow. \label{Nos}
\end{eqnarray}
\noindent Each observed redshift bin contains those galaxies which do not contaminate the other bin (e.g. $\NoT(1-\fow)$) plus those galaxies which are contaminating from the other bin (e.g. $\NwT\fwo$). Note that the total number of galaxies, $\Noo+\Nwo=\NoT+\NwT$, is conserved.

Inverting Eq.~(\ref{eq:wiiomatrix}) allows us to express the true auto-correlations in terms of the observed auto-correlations. The resulting equalities can be plugged into Eq.~(\ref{eq:nbin_wijo}) yielding
\begin{equation}
 \wowo = \frac{\wooo\frac{\Noo}{\Nwo}\fow\fww + \wwwo\frac{\Nwo}{\Noo}\fwo\foo }{(\foo\fww + \fow\fwo)}. \label{eq:wijo_2bin}
\end{equation}

As long as the two bins considered comprise the entire sample of galaxies, this formalism is consistent. Considering a sub-sample allows for leakage to and from the region exterior to the two bins. This can induce a cross-correlation due to mutual contamination of the considered bins from the exterior region and breaks the implicit assumption of galaxy number conservation.

The two-bin case has practical applications such as background selection in cluster lensing, where one seeks a background population that does not share members with the selected foreground cluster.

\begin{figure}
\begin{center}
\includegraphics[scale=0.95,angle=0]{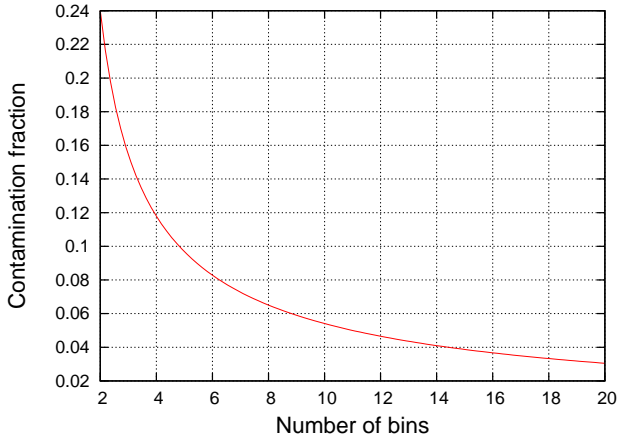}
\caption{\label{fig:table}Upper bounds on the contamination fraction, $f$, as a function of the number of redshift bins n. The upper bound represents the limit at which the pairwise approach breaks down. A global contamination at these levels will cause at most a 10 per cent error in the estimate of $\wiiT$ (Eq.~\ref{eq:wtii_pair}).}
\end{center}
\end{figure}

We now address the global pairwise analysis which --if the contamination fractions are small-- is a good approximation of the full-matrix approach detailed at the beginning of this section. In this case considering each pair of redshift bins in turn will yield estimates for the entire set of contamination fractions. We start by assuming that second order terms in the cross-contaminations are small, such that the off-diagonal elements of the matrix in Eq.~(\ref{eq:wiiomatrix}) satisfy $\fij^2 \ll \fii^2$. This results in the following simple relationship between the observed and true auto-correlations:
\begin{equation}
 \wiiT = \wiio\left(\frac{\Nio}{\NiT}\right)^2\frac{1}{\fii ^2}. \label{eq:wtii_pair}
\end{equation}

When combined with Eq.~(\ref{eq:nbin_wijo}) this yields the following equation for the observed cross-correlation:
\begin{equation}
\wijo=\wiio \frac{\Nio}{\Njo}\frac{\fij}{\fii} + \wjjo \frac{\Njo}{\Nio}\frac{\fji}{\fjj} + \sum^{\rm n}_{\rm k, k \neq i, k \neq j}\wkko\frac{(\Nko)^2}{\Nio\Njo}\frac{\fki\fkj}{(\fkk)^2},\label{eq:pair_cross}
\end{equation}
\noindent All terms that are first order in $f$ have been taken out of the summation operator. This equation is identical to the two bin case of Eq.~(\ref{eq:wijo_2bin}) if we assume that second order terms are negligible.

This approximation will hold as long as the eigenvalues of the n$\times$n matrix of Eq.~(\ref{eq:wiiomatrix}) are well approximated by the diagonal elements. Assuming a uniform cross contamination such that all $\fij$ equal $f$, and therefore $\fii=(1-({\rm n}-1)f)$, we have the following matrix:
\begin{equation}
\begin{pmatrix}
(1-({\rm n}-1)f)^2 & f^2 & \dots & f^2 \\
f^2 & (1-({\rm n}-1)f)^2 & \dots & f^2 \\
\hdotsfor{4} \\
f^2 & f^2 & \dots & (1-({\rm n}-1)f)^2 \\
\end{pmatrix}.
\label{eq:fapprox}
\end{equation}
The eigenvalues are $(1-({\rm n}-1)f)^2 - f^2$ with multiplicity $\rm n-1$ and $(1-({\rm n}-1)f)^2 + ({\rm n}-1)f^2$. The latter eigenvalue, being the larger deviation from the case of a purely diagonal matrix, gives us a constraint on $f$. If we require that the deviation be at most 10 per cent, we find
\begin{equation}
\frac{({\rm n}-1)f^2}{(1-({\rm n} - 1)f)^2} = 0.1,
\end{equation}
\begin{displaymath}
   f = \left\lbrace 
     \begin{array}{lr}
       \frac{1 - (0.1({\rm n}-1))^{-0.5}}{{\rm n}-1-10} & : n \neq 11\\
       (2({\rm n}-1))^{-1} & : n = 11
     \end{array}
   \right.
\end{displaymath}
The limit on the contamination becomes more stringent when more redshift bins are included in the analysis. Figure \ref{fig:table} gives the upper limit on $f$ as a function of the number of redshift bins. The contamination must be less than or equal to these values so that the maximum error made in determining $\wiiT$ via Eq.~(\ref{eq:wtii_pair}) is 10 per cent.

With these limitations in mind we adopt the pairwise analysis throughout the rest of this work. A standard maximum likelihood procedure is used to estimate the contamination fractions by measuring the angular correlation functions at multiple scales and fitting them with Eq.~(\ref{eq:wijo_2bin}). Measurements of the angular correlation functions at different scales are not independent, thus the errors must be described by a covariance matrix. We use a bootstrapping technique to construct the covariance matrix, and include an additional source of error which we refer to as the clustering covariance matrix \citep{2010MNRAS.401.2093V}. We refer the reader to appendix \ref{sec:covariance} for the details of the covariance matrix and the likelihood method.

\section{Application to a simulated galaxy survey}
\label{sec:MilSim}
The Millennium Simulation tracks the hierarchical growth of dark matter structure from a redshift of 127 to the present. The simulation volume is a periodic box of 500 Mpc $\rm h^{-1}$ on a side, containing 2160$^3$ particles each with a mass of $8.6\times10^8 {\rm \Msun}$. The simulation assumes a concordance model $\Lambda$CDM cosmology, $\Omega_{\rm m}=\Omega_{\rm dm} + \Omega_{\rm b}=0.25$, $\Omega_{\rm b}=0.045$, $\Omega_{\Lambda}=0.75$, $h = 0.73$, $n=1$ and $\sigma_{\rm 8}=0.9$, though the details of the cosmology do not affect any of the results presented here. A complete description of the Simulation is presented in \citet{2005Natur.435..629S} and references therein. 

We employ mock observational catalogues constructed from the Millennium Simulation to test the method described in \S\ref{sec:theory}. The catalogues consist of six square pencil beam fields of 1.4 degrees on a side, containing a total of 28.7 million galaxies at redshifts less than 4.0. A full description of the catalogues is given by \citet{2007MNRAS.376....2K}. To identify the mock catalogues, \citet{2007MNRAS.376....2K} give each a label 'a' through 'f'. We maintain this notation throughout the current work when distinguishing between the catalogues.

It is important to note that these mock observational catalogues do not include the effects of lensing, and in particular they do not account for weak lensing magnification. Since magnification will create an angular correlation between high and low redshift, it is relevant to the current work and will be addressed in \S\ref{sec:discussion}.

We first present the results when applying the two-bin analysis to the mock observational catalogues, and discuss the effects of bin width and galaxy density. We then apply the global pairwise analysis demonstrating its ability to recover the true number of galaxies in a redshift bin, and the true average redshift of galaxies in each redshift bin.

\begin{figure*}
\begin{center}
\includegraphics[scale=0.95,angle=0]{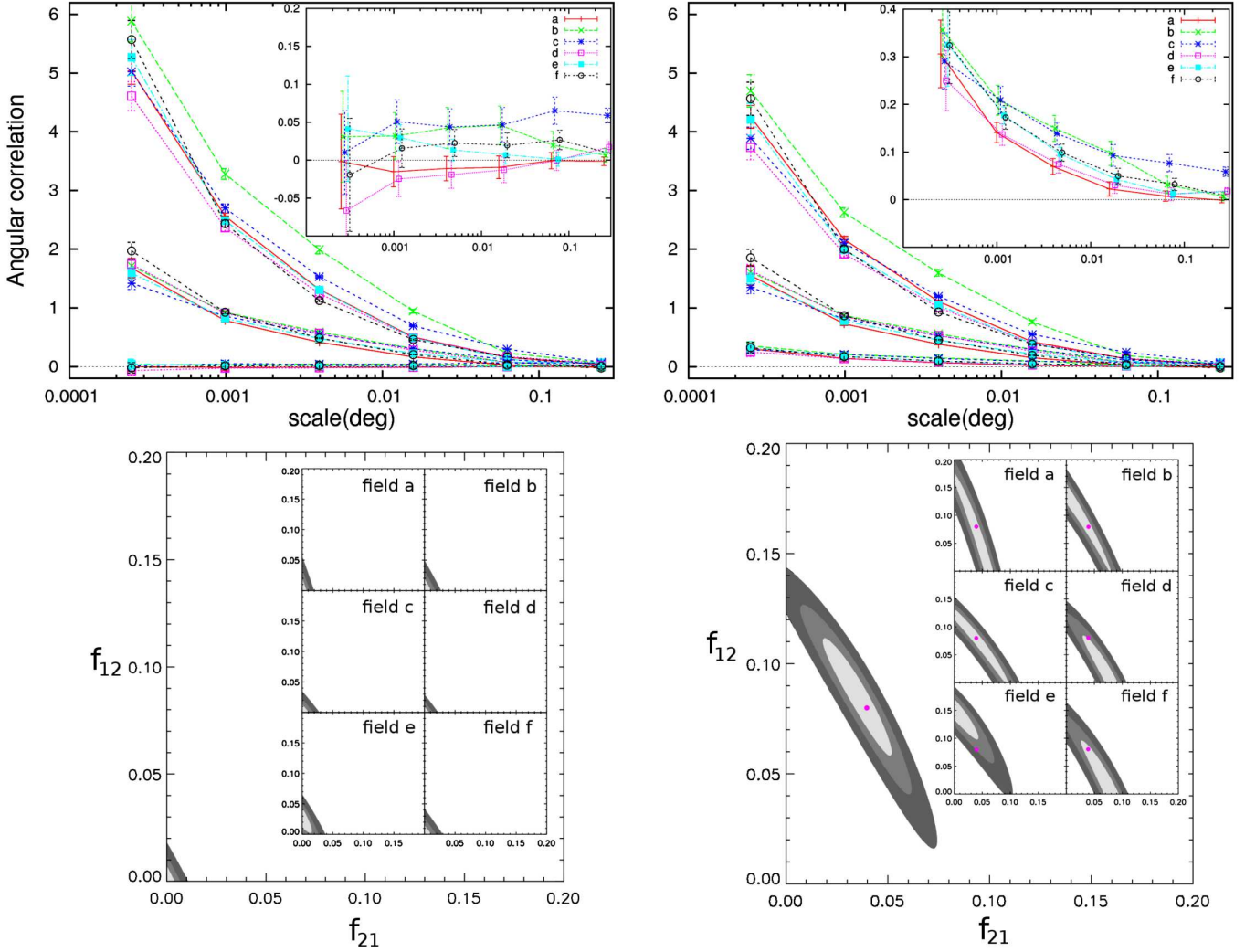}
\caption{\label{fig:one} The left side panels are for the case of no contamination, the right side panels are for a contamination of $\fow=0.08$ and $\fwo=0.04$. \textbf{Top row:} The angular auto- and cross-correlation functions for each test contamination of the mock catalogues. Each of the six fields is plotted with a different line style (colour). The higher redshift bin, $0.8 < z_2 < 0.85$, has a larger amplitude than the lower redshift bin, $0.3 < z_1 < 0.5$. A zoom of the cross-correlation is given in the insert, and the legend therein identifies which line style applies to each of the six mock catalogues. The error bars on the auto-correlation functions come from bootstrapping the catalogue. The cross-correlation functions also include the contribution from the clustering covariance (refer to appendix~\ref{sec:covariance} for details). The cross-correlation is consistent with zero for the case of zero contamination (left panel) and deviates significantly from zero for the contaminated case (right panel) \textbf{Bottom row:} The likelihood region of the contamination fractions. The shaded regions denote the 68, 95 and 99.9 per cent confidence areas, with increasing darkness indicating increasing significance. The contours result from summing the log-Liklihoods for the six individual fields. The insert shows the result for each of the individual fields. The input contamination in the right panel is marked with a dot.}
\end{center}
\end{figure*}

\subsection{Null Test}
\label{sec:NullTest}
Since there is no contamination between redshift bins in the simulated data, the angular cross-correlation function between any two redshift bins ought to be consistent with zero. To test this, two redshift bins are chosen so that there is, approximately, an equal number of galaxies in each; the bins are $0.3 < z_1 < 0.5$ which contains 69,139 galaxies per field on average and $0.8 < z_2 < 0.85$ which contains 50,575 galaxies per field on average. 

\begin{figure}
\begin{center}
\includegraphics[scale=0.85, angle=0]{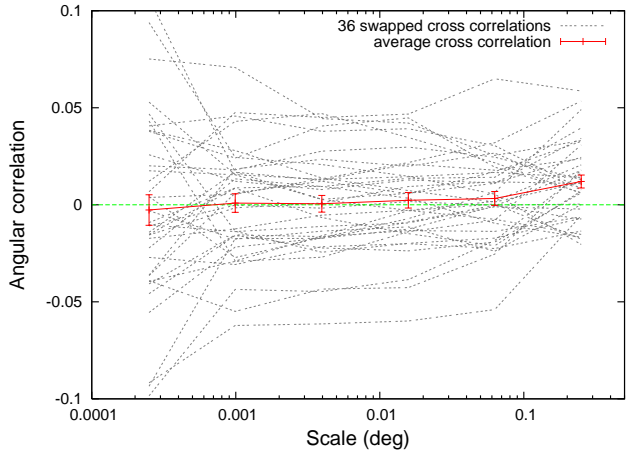}
\caption{\label{fig:swap} The 36 dotted (gray) lines are all of the possible cross-correlations found by swapping the high and low redshift bins between fields. The dashed (green) line shows y=0. The solid (red) line is the average of the 36 cross-correlations, and the error bar is the standard error. The average is consistent with zero, showing that the the clustering term does not bias the cross-correlation.}
\end{center}
\end{figure}

The top left panel of Figure~\ref{fig:one} shows the result of measuring the angular correlation functions on each of the six fields. The auto-correlation functions for a given redshift slice have different amplitudes due to the presence of different structures in each of the fields. The insert shows a zoom of the cross-correlation functions, and the errors include a bootstrapping term as well as the clustering term (see appendix~\ref{sec:covariance}). The clustering term dominates the error at all but the smallest scales, providing a relatively uniform source of noise which behaves like a constant shift in the cross-correlation function. As seen in the the top left panel of Figure~\ref{fig:one}; the six fields provided by the mock catalogues are slightly biased towards a positive shift. However, the clustering term is symmetric about zero and with enough measurements should have no bias. To test this we measured the cross-correlation between all combinations of high- and low-redshift bins, so that the high-redshift bin from field 'a' is cross-correlated with the low redshift bin of fields 'a', 'b', 'c', 'd', 'e' and 'f'. The resulting 36 cross-correlations are displayed in Figure~\ref{fig:swap}, where the solid (red) line shows the average value which is consistent with zero at all scales.

The contamination fractions are estimated from the measured angular correlation functions as described in appendix~\ref{sec:covariance}. The resulting likelihood is presented in the bottom left panel of Figure~\ref{fig:one}. Zero contamination is consistent with the results, and contamination in excess of $\sim$2 per cent can be ruled out at the 99.9 per cent confidence level.

We conclude that, as expected, there is no significant angular cross-correlation between widely separated redshift bins in the Millennium Simulation. Furthermore our method of estimating the contamination performs well in this case, indicating that there is less than $\sim$1 per cent contamination at the 68 per cent confidence level. 

\subsection{Artificial contamination}
Contamination was added to each field by randomly shifting galaxies between the redshift bins. The contamination fractions were taken to be $\fow=0.08$ and $\fwo=0.04$. The measured angular correlations are shown in the top right panel of Figure~\ref{fig:one} where a cross-correlation signal is clearly seen. The recovered likelihood for the contamination fractions is shown in the bottom right panel of Figure~\ref{fig:one}. The input value lies in the 68 per cent confidence region, and zero contamination ($\fow=\fwo=0.00$) is ruled out well beyond the 99.9 per cent confidence level.

The parameter space is highly degenerate; contamination in either direction (low to high redshift, or vice versa) increases the cross-correlation amplitude. Degeneracy breaking comes from the ability to probe the shape of the auto-correlation functions. If only one scale is used, the parameters are completely degenerate. Likewise if the auto-correlation of each redshift bin has the same shape (for example, if they are power laws with the same slope), then the parameters are completely degenerate. 

Each field has independent structure, causing variations in the measured correlation functions. It is important that this is not considered a source of sample variance. Any features in the auto-correlation function will also be present in the cross-correlation function if contamination is present. The inserts in the bottom panels of Figure~\ref{fig:one} show the parameter constraints obtained by analysing each field independently. These contours are combined, yielding a better measurement of the contamination.

We have repeated this procedure for several different contamination fractions ($\fow$,$\fwo$) including ($0.0$,$0.02$), ($0.15$,$0.0$), ($0.0$,$0.15$) and ($0.0$,$0.45$). In all cases the input contamination was recovered within the 68 per cent confidence region. 

\subsection{Effect of galaxy density and redshift bin width}
Each of the six mock catalogues of \citet{2007MNRAS.376....2K} contain on average $4.79\times 10^6$ galaxies out to redshift 4. The two redshift bins used thus far, $0.3 < z < 0.5$ and $0.80 < z < 0.85$, were chosen to be well separated in redshift and contain roughly equal numbers of galaxies. On average there are $\sim 6\times 10^4$ galaxies per redshift bin per field, corresponding to a density of 8.5 arcmin$^{-2}$. 

In order to test the effects of object density, galaxies were randomly removed from the mock catalogues which were then contaminated with $\fow=0.08$ and $\fwo=0.04$. Densities of 4.3 and 1.7 arcmin$^{-2}$ were tested. We found that lower densities yielded only marginally weaker constraints compared to those in the bottom right panel of Figure~\ref{fig:one}.

Reducing object density has a greater effect on constraints from individual fields; much of the constraining power results from the combination of the six fields. As long as there are sufficient galaxies to measure the angular correlation function accurately, the contamination can be constrained. There is a disproportionately small change to the contours when the density is reduced since we are not dominated by Poisson noise.

\begin{figure}
\begin{center}
\includegraphics[scale=0.8,angle=0]{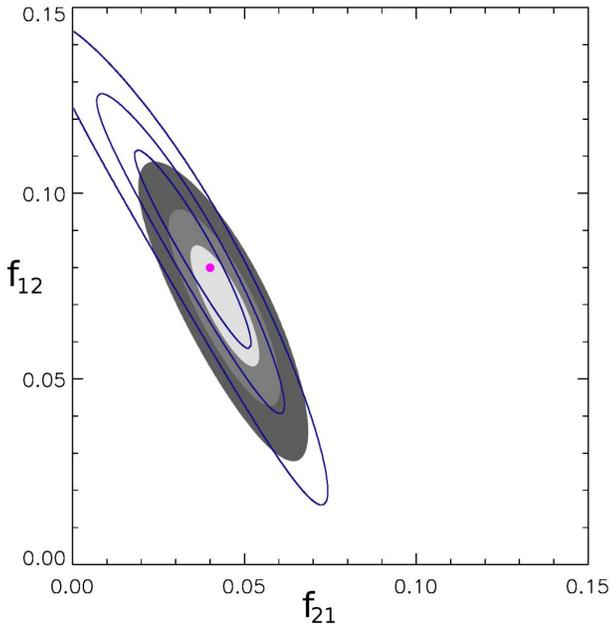}
\caption{\label{fig:smallbin} Constraints on the contamination fractions for an input model of $\fow=0.08$ and $\fwo=0.04$, which is marked with a dot. The lined contours show the results for the wider redshift bins; $0.3 < z < 0.5$ and $0.80 < z < 0.85$. The filled contours give the constraints for more narrow redshift bins; $0.3 < z < 0.4$ and $0.8 < z < 0.82$. Narrow redshift bins provide tighter constraints despite the decreased number density.}
\end{center}
\end{figure}

If more narrow redshift bins are taken, for example $0.3 < z < 0.4$ and $0.8 < z < 0.82$, then the constraints on the contamination improve considerably. Each bin contains 24,451 galaxies per field on average with a corresponding density of 3.5 arcmin$^{-2}$. Figure~\ref{fig:smallbin} shows the parameter constraints when a contamination of $\fow=0.08$ and $\fwo=0.04$ is used, the constraints for wider bins are over-plotted for comparison. The narrow bins offer much tighter constraints despite the density being lowered as a result of the smaller bin size.

The angular correlation function for narrower redshift bins is more distinct since a larger fraction of galaxies are physically clustered with each other. Making the bins wider dilutes the sample with galaxies that are not clustered, reducing the correlation. Arbitrarily narrow bins will cluster strongly, providing a good measure of the correlation function, but will suffer due to low number densities. More work is needed to determine an optimal trade-off between these parameters.

\subsection{Global pairwise analysis}
\label{sec:globalPairwise_MilSim}
We divide the data into the following redshift bins: [0.0,0.5], (0.5,0.8], (0.8,1.1] and (1.1,1.5], which we label z$_1$, z$_2$, z$_3$ and z$_4$ respectively. The average number of galaxies in each bin from low to high redshift is: 60804, 94907, 63707, 55290. In order to test the global pairwise analysis, contamination fractions between all redshift bins must be specified. In lieu of a completely random contamination matrix we have taken one which is similar to the contamination we measure for the CFHTLS-Deep fields in \S~\ref{sec:Deep}. Since the measured contamination in the Deep fields appears to be on the extreme end of the global pairwise approximation we have reduced some of the contamination values, though the adopted contamination matrix remains a strong test of the global pairwise approximation. The contamination matrix used is,
\begin{equation}
\label{eq:input_contam}
\fij=
\begin{pmatrix}
0.70 & 0.01  & 0.10 & 0.10 \\
0.10 & 0.925 & 0.15 & 0.04 \\
0.05 & 0.06  & 0.70 & 0.08 \\
0.15 & 0.005 & 0.05 & 0.78 \\
\end{pmatrix},
\end{equation}
\noindent where $\fij$ refers to the entry in the $\rm i^{th}$ column and the $\rm j^{th}$ row. Galaxies are chosen randomly to be moved between redshift bins such that Eq.~(\ref{eq:Niomatrix}) is satisfied. 

\begin{figure*}
\begin{center}
\hbox{
\includegraphics[scale=0.43,angle=0]{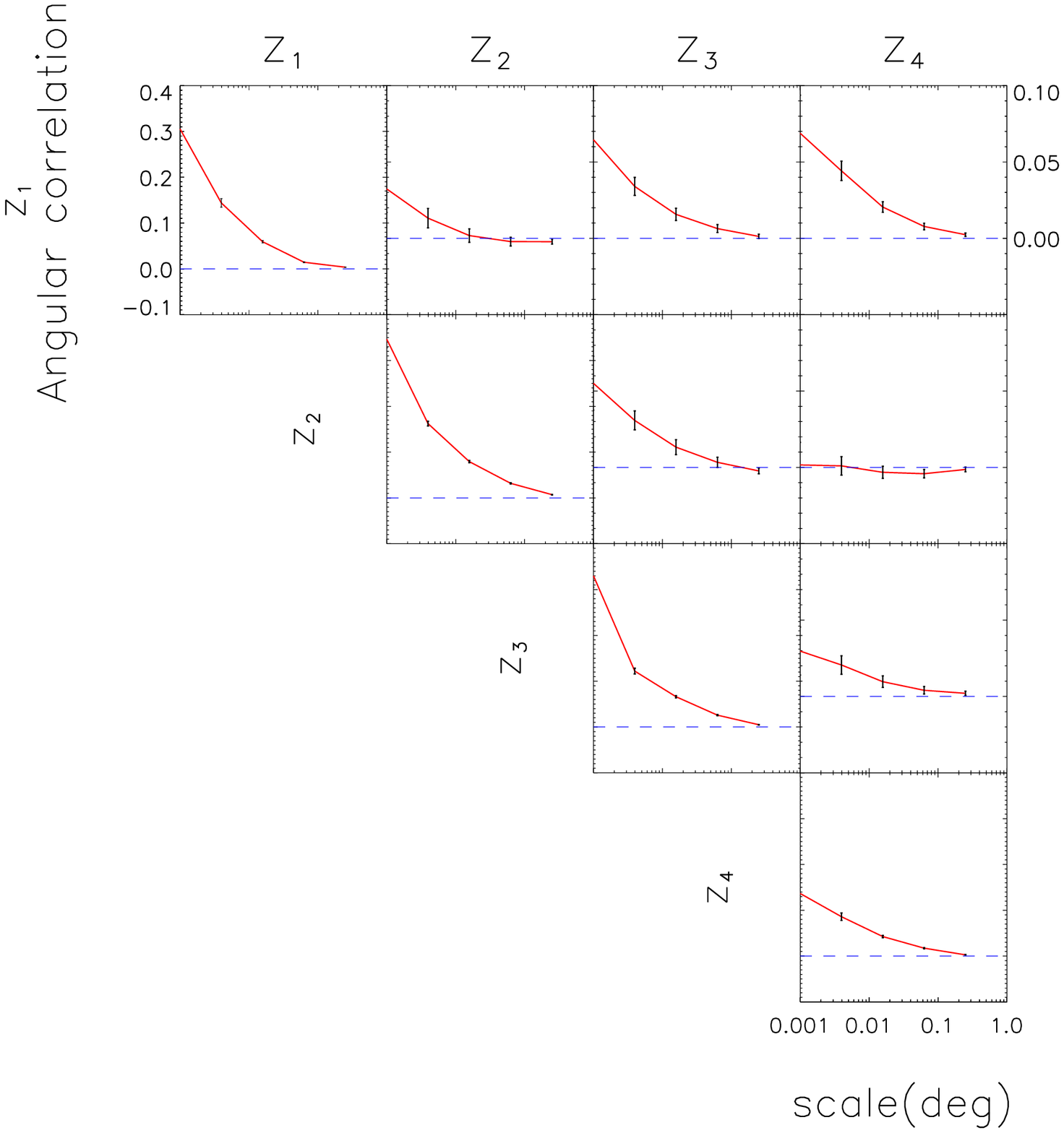}
\includegraphics[scale=0.9,angle=0]{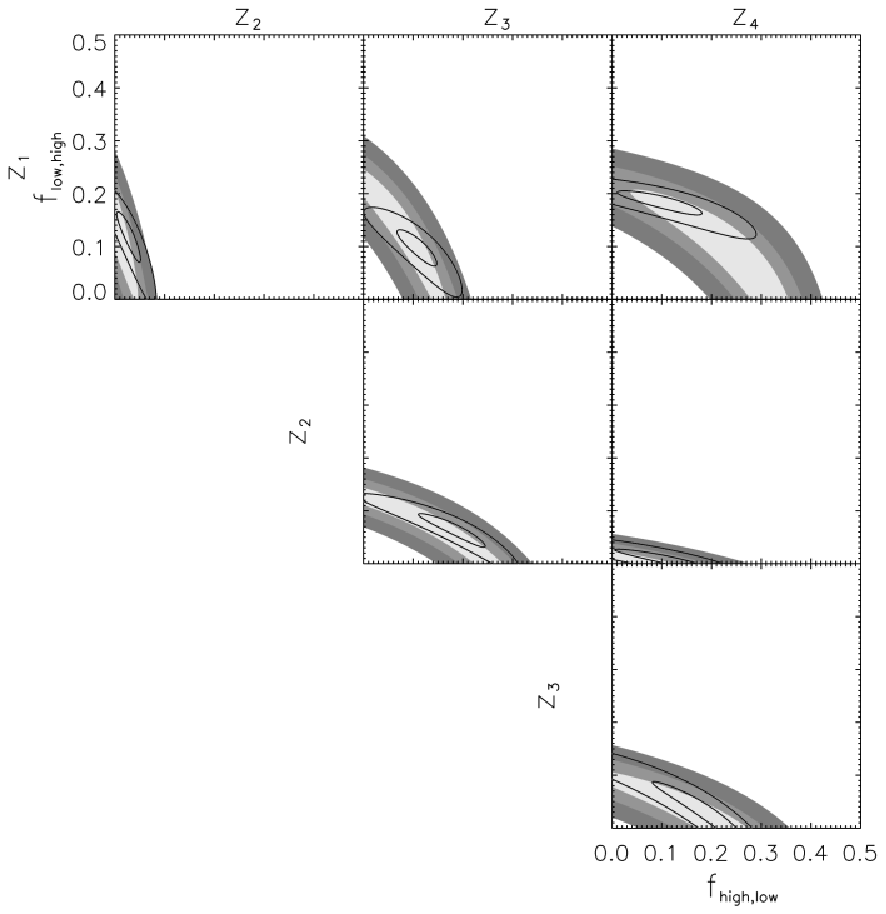}
}
\caption{\label{fig:likeli_MilSim} \textbf{Left panel:} The angular correlation functions as measured for field 'a'. The subplots along the diagonal contain the auto-correlation for each redshift bin. The off-diagonal subplots contain the cross-correlation function between two bins. The redshift bin labels at the top of each column and to the left of each row denote which bins are involved in the given correlation function. The vertical scale on the top-left subplot applies to all subplots on the diagonal, the scale on the top-right subplot applies to all off-diagonal subplots. \textbf{Right panel:} The filled contours depict the parameter constraints estimated for field 'a'. The shading from light to dark represents the 68, 95 and 99.9 per cent confidence levels. The lined contours are the result of combining the parameter constraints from each individual field, for clarity we only show the 68 and 99.9 per cent confidence levels. Each subplot contains the constraints for a pair of contamination fractions. The x-axis is taken to be the contamination fraction from the high redshift bin to the low redshift bin and the y-axis is the reverse. The two bins contributing to each subplot are indicated by redshift bin labels at the top of each column and to the left of each row.}
\end{center}
\end{figure*}

The result of measuring the angular auto-correlation function for each redshift bin and the cross-correlation function for each pair of redshift bins, as measured from field 'a', is presented in the left panel of Figure~\ref{fig:likeli_MilSim}. One hundred bootstraps of the contaminated catalogues were constructed, from which the angular correlation functions were measured. These hundred measurements were used to estimate the covariance matrix, applying the correction described in \citet{Hartlap2007}. For the auto-correlation functions (subplots on the diagonal) this is the only source of error, the cross-correlation functions (off-diagonal subplots) have the clustering term as an additional source of error. 

The pairwise analysis is applied to each pair of redshift bins yielding constraints on the contamination between each pair. The constraints on the contamination fractions for field 'a' are given by the filled contours in the right panel of Figure~\ref{fig:likeli_MilSim}. Each of the six fields is considered independent, and their constraints on the contamination fractions are combined by multiplying the likelihoods together. These combined constraints are given as lined contours, over-plotted in Figure~\ref{fig:likeli_MilSim}.

The pairwise analysis does not impose global constraints since each pair of bins is considered independently. Not all combinations of contamination fractions suggested by Figure~\ref{fig:likeli_MilSim} will produce a consistent picture. For example, it is possible to select points from each plot in the right most column such that more than 100 per cent of galaxies from the highest redshift bin are contaminating other bins. To get a sense of which solutions are globally consistent, and what a typical global solution looks like, we employ a Monte-Carlo method. We first randomly sample 1000 points from each of the pairwise likelihood regions such that the density of points reflects the underlying probability distribution. Taking a random point from each pairwise analysis will uniquely specify a realization of the global contamination, i.e. the matrix of Eq.~(\ref{eq:Niomatrix}). Thus a realization of the true redshift distribution can be determined. We then impose two constraints: the true number of galaxies in a bin cannot be less than zero, and no more than 100 per cent of galaxies can be scattered from a single bin. This procedure is repeated until 200,000 admissible realizations are found. 

We have verified that the admissible realizations of the contamination fractions are representative of the full probability distributions they are drawn from. It is not the case that the global constraints exclude particular regions of parameter space. This is checked by constructing a probability distribution for each contamination fraction from the 200,000 measurements and comparing it to the marginalised PDF obtained from the likelihood contours. In all cases the two agree with each other. For the combined case, the estimated contamination fractions (taken to be the average value) with 68 per cent confidence levels are as follows,
\begin{equation}
\fij=
\begin{pmatrix}
0.60^{+0.04}_{-0.05} & 0.03^{+0.02}_{-0.02} & 0.11^{+0.03}_{-0.03} & 0.11^{+0.05}_{-0.06} \\
0.12^{+0.04}_{-0.03} & 0.89^{+0.04}_{-0.03} & 0.18^{+0.05}_{-0.04} & 0.09^{+0.05}_{-0.05} \\
0.10^{+0.02}_{-0.02} & 0.07^{+0.02}_{-0.02} & 0.66^{+0.06}_{-0.06} & 0.16^{+0.07}_{-0.03} \\
0.18^{+0.02}_{-0.01} & 0.01^{+0.00}_{-0.01} & 0.05^{+0.02}_{-0.05} & 0.65^{+0.09}_{-0.09} \\
\end{pmatrix}.
\end{equation}
\noindent This result agrees extremely well with the input contamination matrix of Eq.~(\ref{eq:input_contam}), with 11 of the 16 contamination fractions, 69 per cent, agreeing within the 68 per cent error estimate. As expected the pairwise analysis slightly overestimates the level of contamination, which is probably more pronounced in this case since the contamination matrix is so aggressive.

\begin{figure}
\begin{center}
\includegraphics[scale=0.28,angle=0]{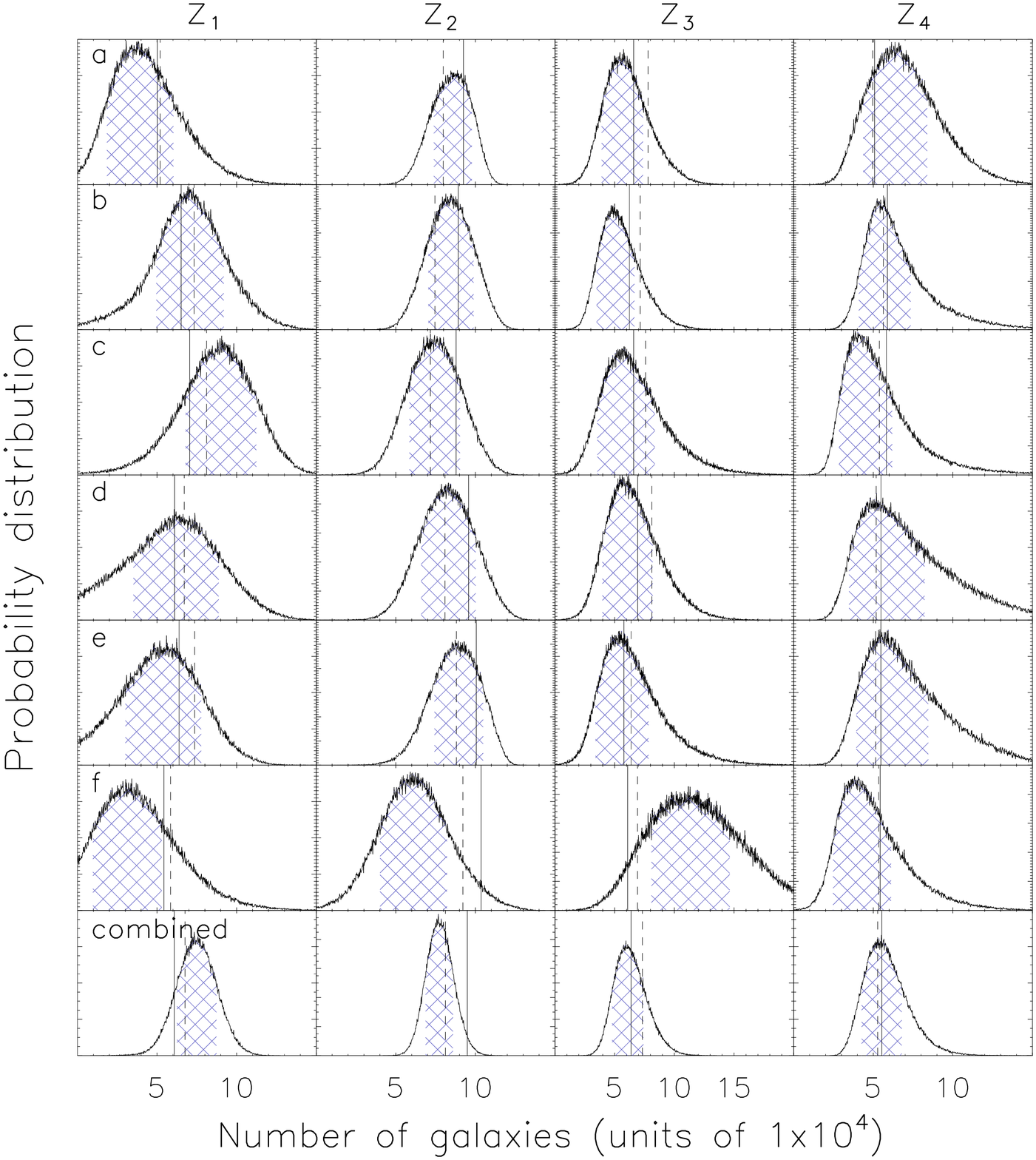}
\caption{\label{fig:nofzPDF_MilSim} This demonstrates the ability of the global pairwise analysis to estimate the true (uncontaminated) redshift distribution. The x-axis is the number of galaxies in units of $1\times10^4$. The y-axis is the probability, which has been scaled differently in each subplot for clarity. The histograms are the probability distribution of the true number of galaxies, and the cross-hashing denotes the 68 per cent confidence region. Each row of subplots is the result from one of the Millennium Simulation fields. The bottom row is the result when the constraints on the contamination fractions for each field are combined (see Figure~\ref{fig:likeli_MilSim}). Please note that the bottom row is not a direct combination of the results from the other rows. Each column represents a redshift bin, as labelled. The solid vertical line in each subplot indicates the observed number of galaxies. The dashed vertical line is the true number of galaxies.}
\end{center}
\end{figure}

Using Eq.~(\ref{eq:Niomatrix}) allows us to estimate the true number of galaxies in each redshift bin for each of the 200,000 realizations of the contamination matrix. This is done for each of the six fields as well as the combined case, the probability distributions for the estimated true number of galaxies are presented in Figure~\ref{fig:nofzPDF_MilSim}. The cross hashed regions denote the 68 per cent confidence level. Since this is simulated data with artificial contamination we know what the actual uncontaminated number of galaxies is for a given bin, this is over-plotted as a dashed vertical line, for clarity we will refer to this as the true number of galaxies, and to our attempt to recover this as the estimated true number of galaxies. The solid vertical line shows the observed number of galaxies; i.e., the number of galaxies after contamination. For the combined case the number of observed galaxies is taken to be the average number of galaxies from the 6 fields. The global pairwise analysis does a good job of recovering the true number of galaxies in each bin, although in many cases the the observed number is also an acceptable fit, owing to the small contamination fractions and the fact that the redshift distribution was nearly flat to begin with. Focusing on the combined result for $z_2$ we see that the true number of galaxies is a significantly better fit to the estimated true number than the observed number.

It is also possible to estimate the true average redshift of a redshift bin. Where the true average redshift is defined as the average of the true redshifts of galaxies in a redshift bin that has contamination. With real world data it is the same as asking what the average spectroscopic redshift is in a given photometric redshift bin. This is straight forward to calculate for a simulated survey since we know for each redshift bin the fraction of galaxies that came from each other redshift bin. The true average redshift is given by:
\begin{equation}
\label{eq:avgz}
\bar{z}_{\rm i}^{\rm T}=\frac{1}{\Nio}\sum_{\rm k=1}^{\rm n}\bar{z}_{\rm k}^{\rm uncontam}\fki\NkT ,
\end{equation}
\noindent where $\bar{z}_{\rm k}^{\rm uncontam}$ is the uncontaminated average redshift of a galaxy in bin k, and $\fki\NkT$ is the number of galaxies in bin i from bin k. In general the uncontaminated average redshift of a bin is not known since it requires knowledge of the true redshift of each galaxy. If we assume that the contamination does not significantly alter the shape of the redshift distribution at the sub-redshift bin level, then the average redshift of a bin will not be changed by contamination. Therefore we estimate the average redshift of an uncontaminated redshift bin ($\bar{z}_{\rm k}^{\rm uncontam}$) by the average redshift of the contaminated redshift bin.

\begin{figure}
\begin{center}
\includegraphics[scale=0.28,angle=0]{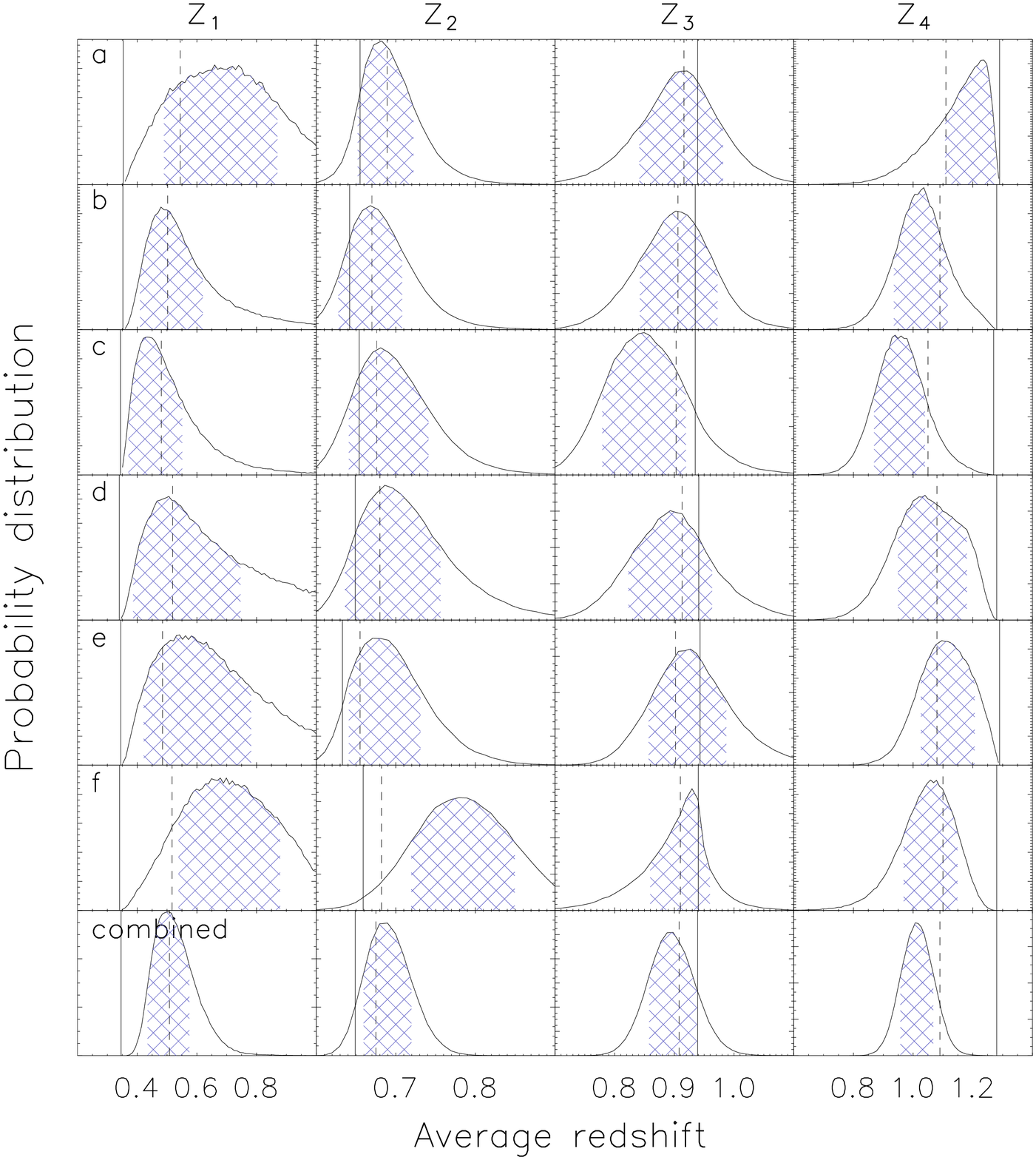}
\caption{\label{fig:avgzPDF_MilSim} The global pairwise analysis is used to estimate the true average redshift of each photometric redshift bin. The x-axis is the average redshift. The y-axis is the probability, which has been scaled differently in each subplot for clarity. The histograms are the probability distribution of the average redshift, and the cross-hashing denotes the 68 per cent confidence region. Each row of subplots is the result from one of the Millennium Simulation fields. The bottom row is the result when the constraints on the contamination fractions for each field are combined (see Figure~\ref{fig:likeli_MilSim}). Please note that the bottom row is not a direct combination of the results from the other rows. Each column represents a redshift bin, as labelled. The solid vertical line in each subplot indicates the average redshift as measured from the uncontaminated catalogue. The dashed vertical line is the true average redshift.}
\end{center}
\end{figure}

For a given field, each of the 200,000 realizations of the contamination matrix yields an estimate of the true average redshift for each redshift bin. Thus a probability distribution function is constructed for each bin and each field, including the combined case, this result is presented in Figure~\ref{fig:avgzPDF_MilSim}. The solid vertical lines show the average redshift with no contamination, this is what one would measure to be the average reshshift if the effects of contamination were ignored. The dashed vertical lines show the true average redshfit for each bin, which we can measure directly here because we're working with simulated data with known contamination. Our method does a very good job of recovering the true average redshift. The lowest and highest redshift bins suffer the most, this is expected since they can only be contaminated by galaxies that are either higher or lower in redshift respectively, whereas the middle bins are contaminated by galaxies which are both higher and lower than the average allowing for a cancellation of the effect.

We have demonstrated that the global pairwise analysis can be used to reliably recover small contaminations between redshift bins. Using the Monte Carlo approach detailed in this section we have shown that we are able to estimate the true redshift distribution, and the true average redshift within a bin. 

\section{Application to a real galaxy survey}
\label{sec:Deep}
The Canada-France-Hawaii Telescope Legacy Survey (CFHTLS) is a joint Canadian-French program designed to take advantage of Megaprime, the CFHT wide-field imager.  This 36-CCD mosaic camera has a $1 \times 1$ degree field of view and a pixel scale of 0.187 arcseconds per pixel. The deep component consists of four one-square-degree fields imaged with five broad-band filters: $u^*$, $g'$, $r'$, $i'$, $z'$. The fields are designated D1, D2, D3 and D4, and are centred on RA;DEC coordinates of 02 26 00; -04 30 00, 10 00 28; +02 12 21, 14 19 28; +52 40 41 and 22 15 32; +17 44 06 respectively

We use the deep photometric redshift catalogues from \citet{Ilbert..0603217} who have estimated redshifts for the T0003 CFHTLS-Deep release.  The redshift catalogue is publicly available at $\it{terapix.iap.fr}$. The full photometric catalogue contains 522286 objects, covering an effective area of 3.2 deg$^2$. A set of 3241 spectroscopic redshifts with $0 \leq z \leq 5$ from the VIRMOS VLT Deep Survey (VVDS) were used as a calibration and training set for the photometric redshifts. The resulting photometric redshifts have an accuracy of $\sigma_{(z_{phot} - z_{spec})/(1+z_\mathrm{spec})} = 0.043$ for $i'_{\mathrm{AB}}$ = 22.5 - 24, with a fraction of catastrophic errors of 5.4 per cent. 

\begin{figure}
\begin{center}
\includegraphics[scale=0.89,angle=0]{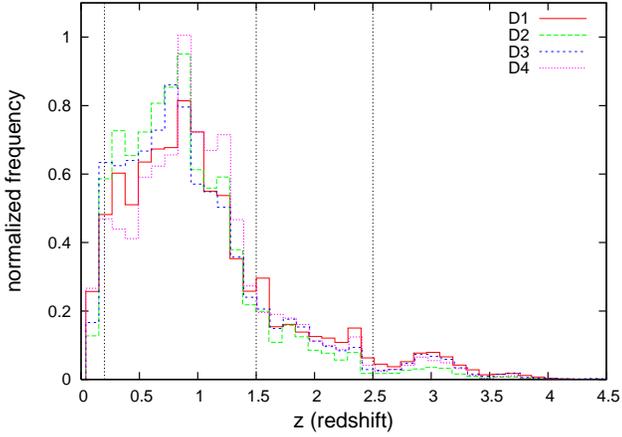}
\caption{\label{fig:nofzDeep} Finely binned redshift distribution for each of the four deep fields. The vertical lines denote the binning adopted in our global pairwise analysis. Note the bump near redshift 3.}
\end{center}
\end{figure}

In this work we consider galaxies with $21.0 < i' < 25.0$ and divide the data into the following redshift bins: [0.0,0.2], (0.2,1.5], (1.5,2.5] and (2.5,6.0], which we label z$_1$, z$_2$, z$_3$ and z$_4$ respectively. The average number of galaxies in each bin from low to high redshift is: 5772, 96019, 15546, 5315. These redshift bins are chosen to isolate the low confidence redshifts and to probe the bump in the redshift distribution near redshift 3 (see Figure~\ref{fig:nofzDeep}). Adopting the definition of catastrophic error used by \citet{Ilbert..0603217}, $\Delta z > 0.15(1+z)$, we measure the fraction of galaxies with catastrophic errors in each bin. From low to high redshift we find: 23.2, 12.4, 33.2 and 23.1 per cent. It is difficult to interpret these values in the context of the contamination fractions. Photometric redshifts with large errors do not necessarily contaminate other redshift bins. Furthermore $\Delta z$ does not take into account degeneracies in the spectral template fitting that allow for alternative solutions to the galaxy's redshift. The catastrophic errors give some indication of the leakage between adjacent redshift bins but cannot account for those galaxies which are completely misclassified.

\subsection{Applying the global pairwise analysis}
\begin{figure*}
\begin{center}
\hbox{
\includegraphics[scale=0.43,angle=0]{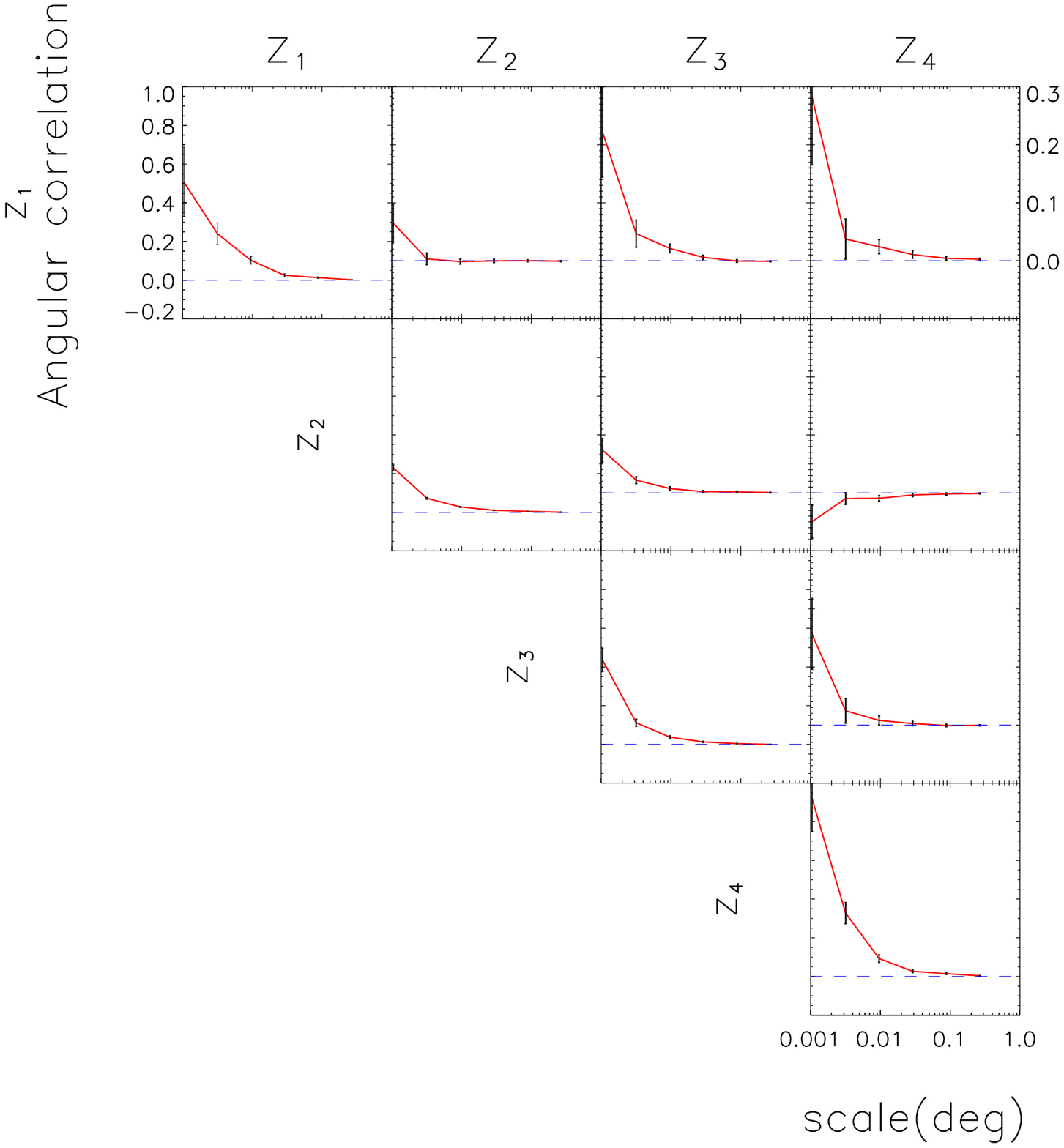}
\includegraphics[scale=0.9,angle=0]{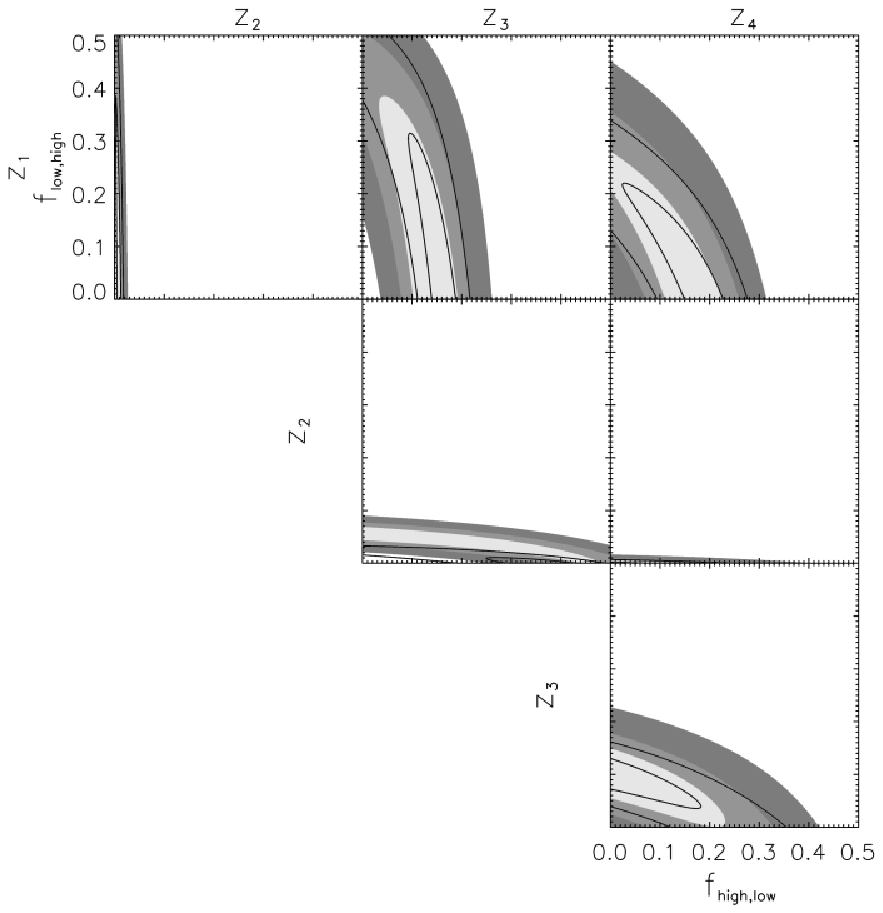}
}
\caption{\label{fig:likeli_D1} Same as caption to Figure~\ref{fig:likeli_MilSim} but with field D1 in place of field 'a'.}
\end{center}
\end{figure*}

We apply the two-bin analysis between each pair of redshift bins, and for each of the four Deep fields. The measured angular correlation functions for D1 are presented in Figure~\ref{fig:likeli_D1}. The covariance matrices are estimated by bootstrapping the catalogues 100 times, and applying the correction described by \citet{Hartlap2007}. For the cross-correlation covariance matrix we also calculate the clustering covariance described in appendix~\ref{sec:covariance}. Parameter constraints are estimated for each cross-correlation, those for D1 are presented in Figure~\ref{fig:likeli_D1}. The other three fields have very similar angular correlation functions and parameter constraints. The parameter constraints for all of the fields can be combined by treating them as statistically independent and multiplying their likelihoods together, yielding tighter constraints (lined contours in Figure~\ref{fig:likeli_D1}).

We constructed 200,000 realizations of a globally consistent contamination matrix, as detailed in \S\ref{sec:globalPairwise_MilSim}, and verified that the admissible realizations are representative of the full probability distributions they are drawn from. The contamination matrix estimated from the combined constraints is,
\begin{equation}
\fij=
\begin{pmatrix}
0.50^{+0.17}_{-0.17} & 0.006^{+0.004}_{-0.003} & 0.13^{+0.04}_{-0.02} & 0.13^{+0.07}_{-0.04} \\
0.18^{+0.06}_{-0.18} & 0.979^{+0.014}_{-0.006} & 0.31^{+0.08}_{-0.04} & 0.04^{+0.01}_{-0.04} \\
0.16^{+0.05}_{-0.16} & 0.006^{+0.001}_{-0.006} & 0.53^{+0.05}_{-0.10} & 0.09^{+0.03}_{-0.09} \\
0.11^{+0.03}_{-0.11} & 0.001^{+0.000}_{-0.001} & 0.07^{+0.04}_{-0.02} & 0.74^{+0.10}_{-0.10} \\
\end{pmatrix},
\end{equation}
\noindent where the entries represent the average value calculated from their probability distribution. The maximum contamination fraction for four redshift bins presented in Figure~\ref{fig:table} is $\sim$0.12 which is smaller than some of the contamination fractions found here (or seen in Figure~\ref{fig:likeli_D1}). It is possible that this is an indication that the pairwise analysis does not hold for these data; however, several assumptions were made in deriving the maximum contamination fraction which do not hold here. We assumed that all contamination fractions have the same value; this is not the case here, and although some tend to be larger than 0.12, some are very close to zero, and several have large errors encompassing zero. We also assumed that there were equal numbers of galaxies in each bin. The ratio of the number of galaxies between the pair of bins enters into the expression for the observed cross-correlation. Since the first and last redshift bins have far fewer galaxies, a large contamination fraction from one of these bins represents only a small number of galaxies. We have also demonstrated our ability to recover the contamination fractions from a similarly aggressive contamination matrix in \S\ref{sec:globalPairwise_MilSim}. For these reasons we believe that the global pairwise analysis remains a good approximation here.

\begin{figure}
\begin{center}
\includegraphics[scale=0.28,angle=0]{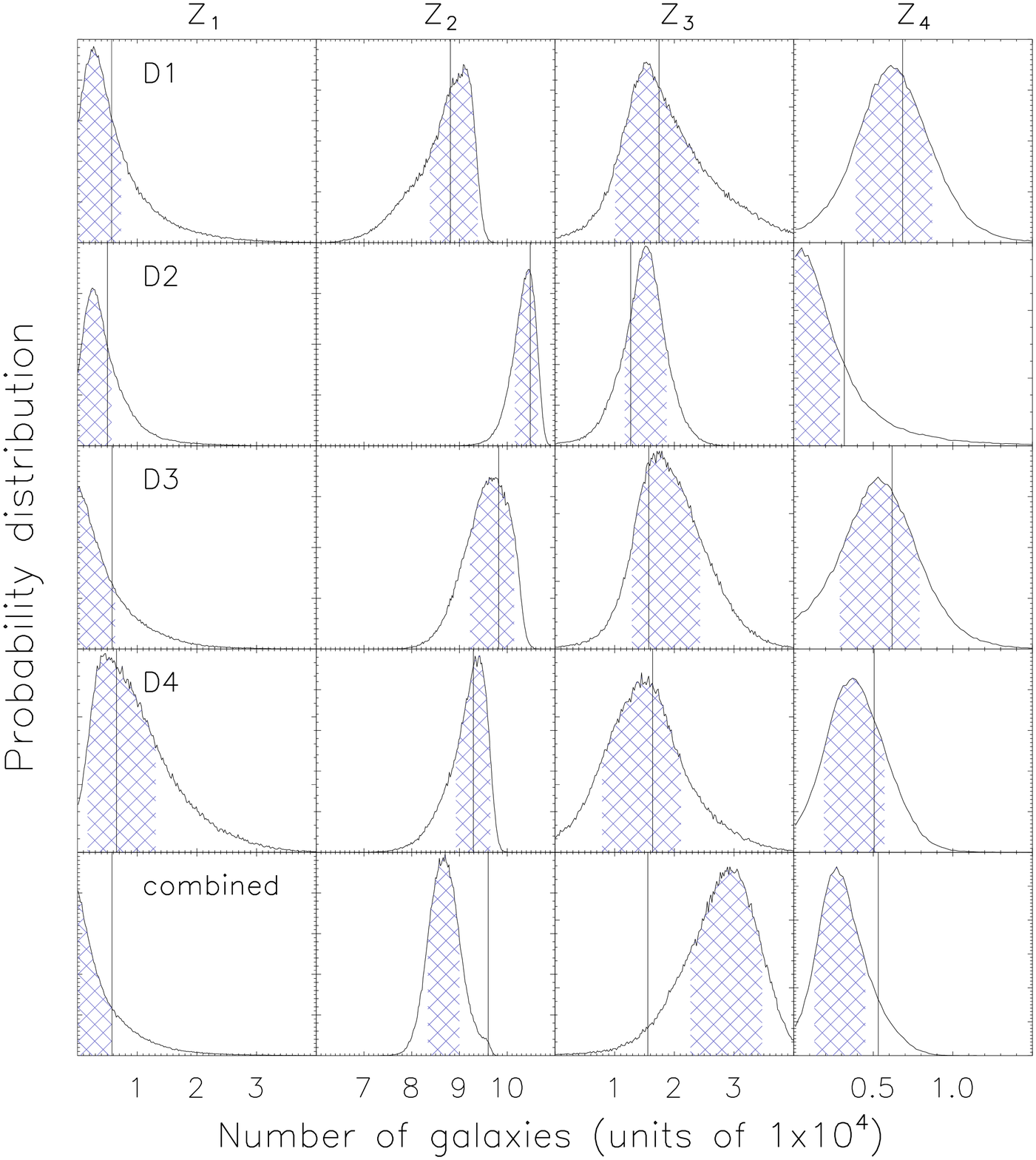}
\caption{\label{fig:nofzPDF} This demonstrates the ability of the global pairwise analysis to reconstruct the true (uncontaminated) redshift distribution. The x-axis is the number of galaxies in units of $1\times10^4$. The y-axis is the probability, which has been scaled differently in each subplot for clarity. The histograms are the probability distribution of the true number of galaxies, and the cross-hashing denotes the 68 per cent confidence region. Each row of subplots is the result from one of the CFHTLS-Deep fields. The bottom row is the result when the constraints on the contamination fractions for each field are combined. Each column represents a redshift bin, as labelled. The vertical line in each subplot indicates the observed number of galaxies.}
\end{center}
\end{figure}

The probability distribution of the true number of galaxies for each redshift bin and each field is presented in Figure~\ref{fig:nofzPDF}; the cross-hatched regions indicate 68 per cent confidence. The observed number of galaxies in each bin is denoted by a vertical line. The bottom row contains the result when the constraints on the contamination fractions for each of the four fields are combined. The smallest fractional change is for $\rm z_2$ which is the high confidence photometric redshift bin. Taking the peak of the probability distribution indicates about a factor of two less galaxies in the highest redshift bin than are observed, suggesting that the bump seen in the photometric redshift distribution is an artefact of contamination. However, with only four square degrees of data, we are unable to rule out the existence of this feature.

The set of globally consistent realizations of the contamination can also be used to estimate the average redshift for each photometric redshift bin. We use Eq.~(\ref{eq:avgz}), and estimate the uncontaminated average redshift of each bin ($\bar{z}_{\rm k}^{\rm uncontam}$) by the average of the photometric redshifts. Which is a good approximation as long as the shape of the observed redshift distribution within each bin is similar to that of the true redshift distribution.

\begin{figure}
\begin{center}
\includegraphics[scale=0.28,angle=0]{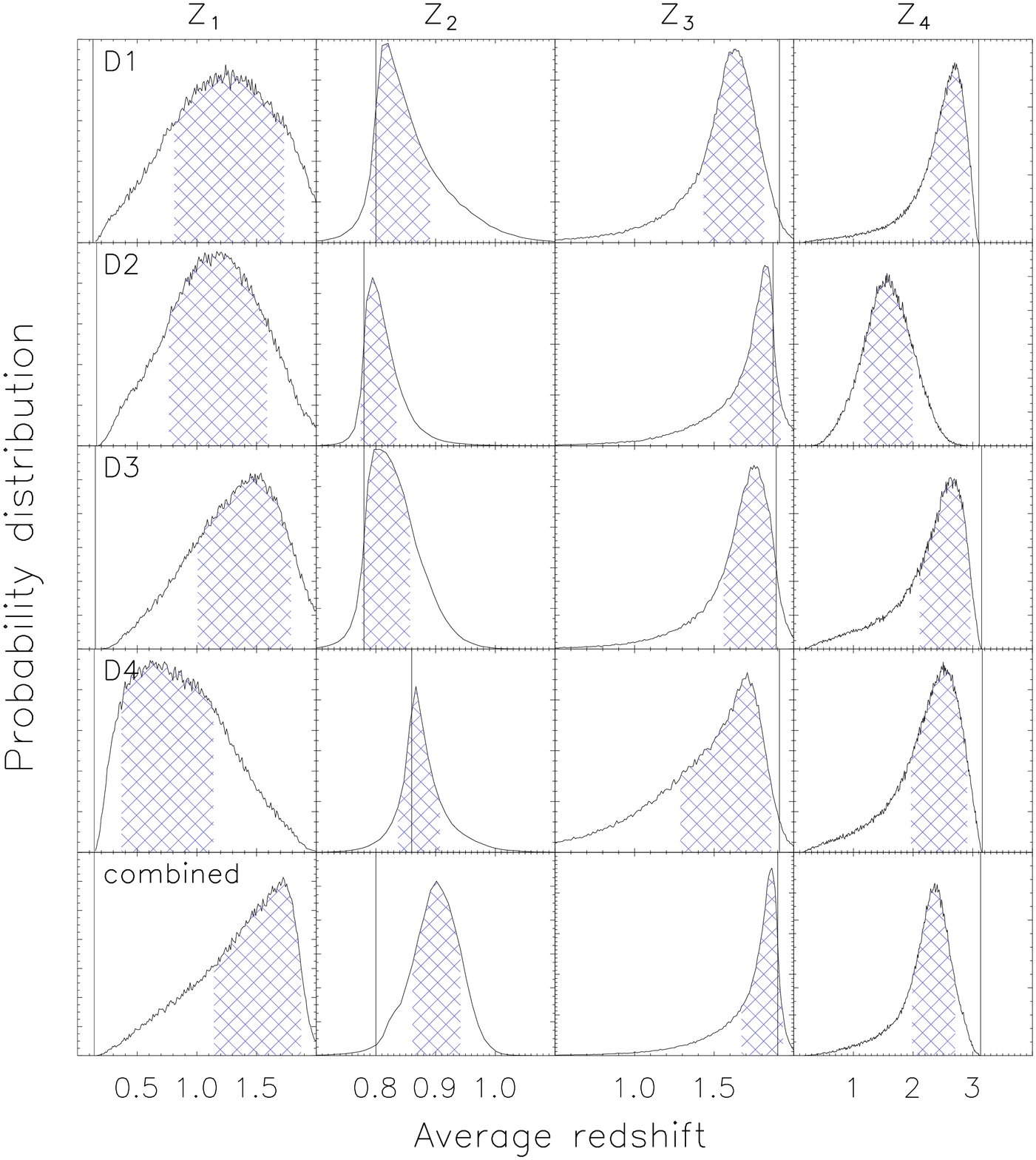}
\caption{\label{fig:avgzPDF} The global pairwise analysis is used to estimate the true average redshift of each photometric redshift bin. The x-axis is the average redshift. The y-axis is the probability, which has been scaled differently in each subplot for clarity. The histograms are the probability distribution of the average redshift, and the cross-hashing denotes the 68 per cent confidence region. Each row of subplots is the result from one of the CFHTLS-Deep fields. The bottom row is the result when the constraints on the contamination fractions for each field are combined. Each column represents a redshift bin, as labelled. The vertical line in each subplot indicates the average redshift as measured from the photometric redshift catalogue.}
\end{center}
\end{figure}

The results are presented in Figure~\ref{fig:avgzPDF} which shows the probability distribution of the average redshift for each redshift bin and each field. Vertical lines show the average redshift for each bin measured from the photometric catalogue. It is clear that the smallest, and largest, redshift bins ($\rm z_1$, and $\rm z_4$) contain galaxies whose true average redshifts deviate significantly from the average redshift expected for those redshift bins. This suggests that many galaxies in bin $\rm z_1$ are in fact from much higher redshifts. Similarly galaxies in bin $\rm z_4$ have a lower than expected average redshift.

\section{Conclusion and Discussion}
\label{sec:discussion}
We have presented an analytic framework for estimating contamination between photometric redshift bins, without the need for any spectroscopic data beyond those used to train the photometric redshift code. To measure the contamination between redshift bins we exploit the fact that mixing between bins will result in a non-zero angular cross-correlation between those bins. We have shown how the contamination will affect the observed angular correlation functions for the general case of contamination between an arbitrary number of bins. For the case of two- and three-bins we explicitly work out the equations.

The case of two-bins is given special attention since it is the simplest case. We note that if the contamination between bins is small enough, then each pair of bins can be considered independently, yielding an accurate measure of contamination between all bins. We refer to this as a global pairwise analysis. 

We test our formalism with mock galaxy catalogues created from the Millennium Simulation. We verify that there is no evidence of contamination, finding an upper limit of $\sim$2 per cent at the 99.9 per cent confidence level. The catalogues are then contaminated by moving galaxies between redshift bins. We demonstrate that the two-bin analysis is able to recover input contamination between redshift bins. The effects of galaxy density and bin-width are investigated. We find that our ability to constrain the contamination fractions is not very sensitive to object density, whereas narrower bins offer better constraints. 

We split the mock catalogues into four redshift bins and apply artificial contamination between all pairs. The global pairwise analysis is used to constrain the contamination fractions between all pairs of redshift bins. A Monte-Carlo method is then used to estimate the true (uncontaminated) redshift distribution, and the true average redshift of galaxies in each bin. This is valuable information for the cosmological interpretation of galactic surveys, and in particular weak lensing by large scale structure. We demonstrate the ability of the method to accurately recover the input contamination as well as reconstruct the true redshift distribution and average redshift of each bin.

The formalism is applied to a real galaxy survey; the four square degree deep component of the Canada-France-Hawaii Telescope Legacy Survey for which there are photometric redshift catalogues \citep{Ilbert..0603217}. We divide the data into four redshift bins and apply the global pairwise analysis. This yields constraints on the contamination fractions, the true redshift distribution, and the true average redshift of galaxies in each bin. We demonstrate here the feasibility of the method with only four square degrees of sky coverage; future application to large galaxy surveys will significantly improve constraining power.

This work has focused on the application of the two-bin and global pairwise methods. For a small number of redshift bins, with sufficiently small contamination, the global pairwise analysis offers a quick and easy means of assessing the contamination between redshift bins. The benefit is largely computational, it is very fast to constrain a model with only two free parameters. A more sophisticated method (such as a Monte-Carlo-Markov-Chain (MCMC)) will be needed to implement the full multi-bin approach. With only three redshift bins there is a total of six free parameters which already renders the simple maximum-likelihood approach impractical. Although the full multi-bin approach will yield the most accurate results, for some applications the more simplistic pairwise analysis should suffice.

Future work will need to take into account the effect of weak lensing magnification, which causes an angular cross-correlation between background galaxies and foreground lenses. Since foreground lenses boost the magnitude of background galaxies there are more close pairs detected between these redshift slices then one would expect from random placement. This effect is well understood and can be easily modelled and accounted for \citep{2005ApJ...633..589S}, however, since it depends on cosmology and the redshifts of the lens and background galaxies it can not be removed in a model-independent way. Dust extinction of the background sources is also important, reducing the brightness of lensed galaxies \citep{2009arXiv0902.4240M}. This de-magnification is wavelength dependant, and in visible passbands is comparable in magnitude to lensing magnification, therefore, it will need to be accounted for along with magnification. 

The expected amplitude of the angular cross-correlation due to magnification is small. Using the CFHTLS-Deep fields, and a magnitude cut similar to that used in this work, \citet{2010MNRAS.401.2093V} find the amplitude of the angular cross-correlation between the redshift bins $z=[0.1,0.6]$ and $z=[1.1,1.4]$ to be about 0.01 on scales smaller than 1 arcmin. While this is clearly an important contribution to the angular cross-correlations measured here it can only account for about 10 per cent of the observed signal on these scales.

Photometric redshifts are more secure for red galaxy types. Furthermore red and blue galaxies cluster differently resulting in distinct angular correlation functions. Therefore, it is likely the case that galaxies doing the contamination are predominantly blue and exhibit a systematically different angular correlation than the red galaxies which do not contaminate. More work is needed to understand the severity of this bias. However, since any cross-correlation signal (above that expected from magnification) indicates contamination it is always possible to use this technique as a null test.

We have presented a method of measuring contamination between photometric redshift bins using the angular correlation function, and without any need for spectroscopically determined redshifts. The method is able to constrain the true redshift distribution and the true average redshift in a photometric bin, both of which are of keen interest to cosmological use of these data. The accuracy of this method will need to be improved to address the needs of high precision cosmology. The inclusion of the galaxy-shear correlation function to break parameter degeneracies has been investigated by \citet{Zhang2009}, showing that the stringent requirements of future surveys can be reached if this information is included. Without the need for accurate weak lensing shear measurements, the method we present here is more accessible and provides valuable information.

\section*{acknowledgments}
JB is supported by the Natural Sciences and Engineering Research Council (NSERC), and the Canadian Institute for Advanced Research (CIAR). LVW is supported by NSERC, CIAR and the Canadian Foundation for Innovation (CFI). The Millennium Simulation databases used in this paper and the web application providing online access to them were constructed as part of the activities of the German Astrophysical Virtual Observatory. This work is partly based on observations obtained with MegaPrime equipped with MegaCam, a joint project of CFHT and CEA/DAPNIA, at the Canada-France-Hawaii Telescope (CFHT) which is operated by the National Research Council (NRC) of Canada, the Institut National des Science de l'Univers of the Centre National de la Recherche Scientifique (CNRS) of France, and the University of Hawaii. This work is based in part on data products produced at
TERAPIX and the Canadian Astronomy Data Centre as part of the Canada-France-Hawaii Telescope Legacy Survey, a collaborative project of NRC and CNRS. This paper makes use of photometric redshifts produced jointly by Terapix and VVDS teams.

\bibliographystyle{benjamin_bibsty}

\appendix
\section{Covariance and likelihood}
\label{sec:covariance}
Here we present the details of the maximum-likelihood method, and the covariance matrix used. Throughout the paper we fit the observed angular cross-correlation between two redshift bins with the model described by Eq.(\ref{eq:wijo_2bin}). Since there are observational errors associated with the angular auto- and cross-correlation functions, we have grouped these quantities on the left hand side of Eq.(\ref{eq:wijo_2bin}), yielding:

\begin{equation}
 \wijo({\vc \theta})(\fii\fjj + \fij\fji) - \wiio({\vc \theta})\frac{\Nio}{\Njo}\fij\fjj - \wjjo({\vc \theta})\frac{\Njo}{\Nio}\fji\fii = 0, \label{eq:wijo_2bin_lhs}
\end{equation}
\noindent where the angular correlation functions are written as a function of scale ${\vc \theta}$. For simplicity let $\Fg$ represent the left hand side of the equation. We therefore seek to calculate the likelihood,

\begin{equation}
{\cal L}={\frac{1}{\sqrt{(2\pi)^s|\Cg|}}} \exp\left[-\frac{1}{2}(\Fg-\mg)\Cg^{-1}(\Fg-\mg)^T\right], \label{eq:likelihood}
\end{equation}
\noindent where $s$ is the number of angular scale bins, $\mg$ is the model which is zero for all scales and $\Cg$ is the $s\times s$ covariance matrix. The covariance matrix is 

\begin{equation}
C_{kl}=\langle \F_k \F_l\rangle,
\label{eq:covsmall}
\end{equation}
\noindent where $k$ and $l$ denote the scales at which the angular correlation functions are measured. Expanding the above yields

\begin{eqnarray}
\label{eq:covlarge}
C_{kl}&=&\langle \wijo(\theta_k)\wijo(\theta_l)\rangle(\fii\fjj + \fij\fji)^2 \\ \nonumber
&+& \langle \wiio(\theta_k)\wiio(\theta_l)\rangle \left(\frac{\Nio}{\Njo}\fij\fjj \right)^2 \\ \nonumber
&+& \langle \wjjo(\theta_k)\wjjo(\theta_l)\rangle \left(\frac{\Njo}{\Nio}\fji\fii \right)^2 \\ \nonumber
&+& \langle \wijo(\theta_k)\wiio(\theta_l)\rangle 2(\fii\fjj + \fij\fji)\frac{\Nio}{\Njo}\fij\fjj \\ \nonumber
&+& \langle \wijo(\theta_k)\wjjo(\theta_l)\rangle 2(\fii\fjj + \fij\fji)\frac{\Njo}{\Nio}\fji\fii \\ \nonumber
&+& \langle \wiio(\theta_k)\wjjo(\theta_l)\rangle 2\fij\fjj\fji\fii. \nonumber
\end{eqnarray}
Ideally the covariance matrix can be estimated directly from the data, but this requires many fields. It is not possible to do this for either the Millennium Simulation or the CFHTLS-Deep data sets which we consider in this work. An alternative is to use a bootstrapping method, wherein the data catalogue is resampled multiple times, and each resampled catalogue is used to measure the angular correlation functions. These angular correlation functions can then be used to calculate the covariance.

\begin{figure}
\begin{center}
\includegraphics[scale=0.26,angle=0]{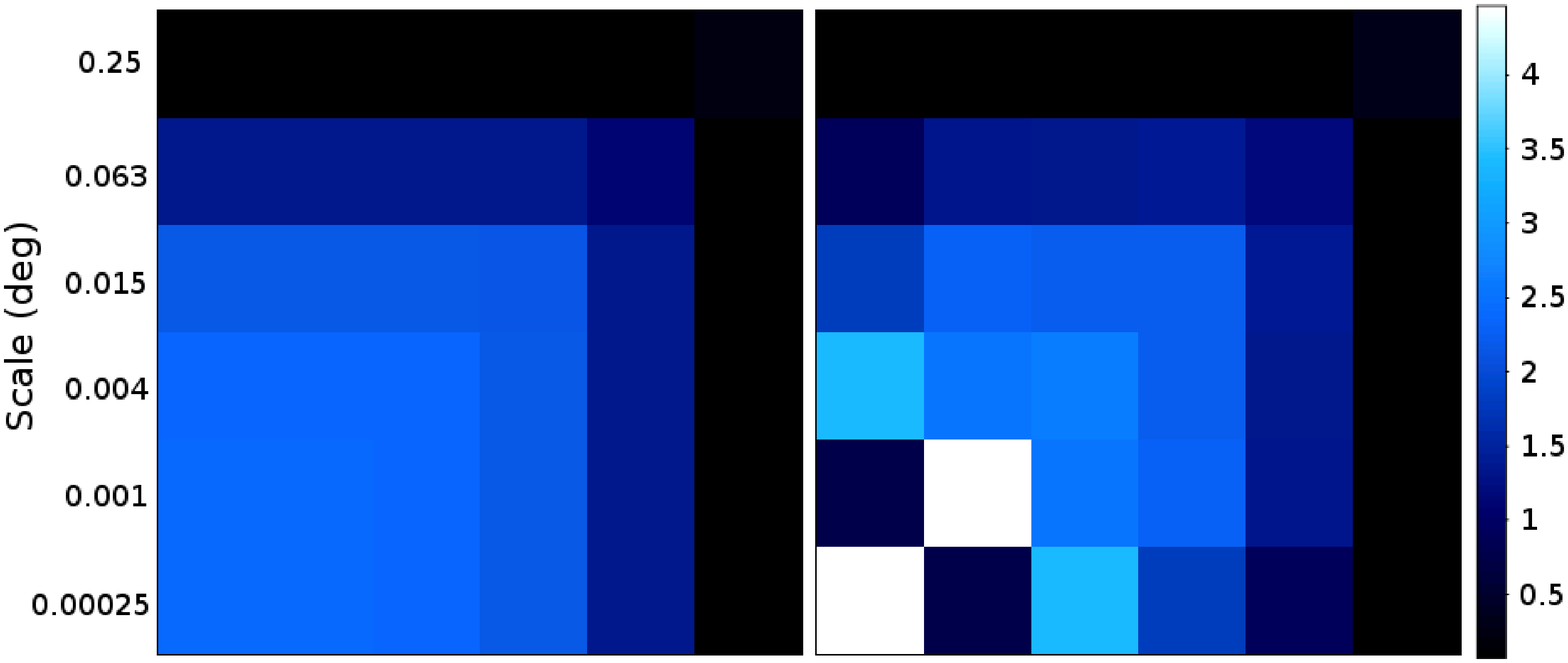}
\caption{\label{fig:covariance} The covariance matrix of the cross-correlation $\langle \wijo(\theta_k)\wijo(\theta_l)\rangle$ has two contributions. We take field 'a' of the Millennium Simulation and two redshift bins as described in \S\ref{sec:NullTest}. The LEFT panel shows the clustering term calculated as described in \citet{2010MNRAS.401.2093V}. The matrices are 6x6 and increase in scale from bottom to top and left to right. The RIGHT panel is the total covariance which includes the bootstrap covariance matrix in addition to the clustering covariance. The scale on the right is in units of $10^{-4}$. Note that the clustering term results in a very flat covariance between all scales, whereas the bootstrap covariance adds relatively little and only on the smallest scales.}
\end{center}
\end{figure}

This procedure suffices for all contributions to the covariance matrix Eq.~(\ref{eq:covlarge}), except the first term $\langle \wijo(\theta_k)\wijo(\theta_l)\rangle$. As shown in \citet{2010MNRAS.401.2093V}, the covariance of the cross-correlation function has a contribution due to the intrinsic clustering of the background and foreground populations. This so-called clustering term can be calculated analytically given the auto-correlation functions and the survey geometry. We add the covariance due to the clustering term to our bootstrap covariance matrix. The left panel in Figure~\ref{fig:covariance} shows the covariance from the clustering term and the right panel shows the total covariance of the cross-correlation function between two redshift bins of the Millennium Simulation (see \S\ref{sec:NullTest}). This constitutes the largest contribution to the covariance matrix of Eq.~(\ref{eq:covlarge}).

\onecolumn
\section{Solving the three-bin case analytically}
\label{sec:threebin}
For three bins it is easy to derive equations for the three observed cross-correlation functions, analogous to what is presented in Eq.~(\ref{eq:wijo_2bin}) for the two-bin case. Let the n$\times$n matrix in Eq.~(\ref{eq:wiiomatrix}) be called F. The inverse can be calculated using the adjoint,
\begin{equation}
 \rm{F}^{-1}=\frac{\rm{adj(F)}}{\rm{det(F)}}.\label{eq:inverse}
\end{equation}
\noindent Both the adjoint (adj(F)) and the determinant (det(F)) can be calculated easily:
\begin{eqnarray}
\rm{adj(F)}&=& 
\begin{pmatrix}
  f_{\rm 22}^2f_{\rm 33}^2-f_{\rm 32}^2f_{\rm 23}^2 & f_{\rm 31}^2f_{\rm 23}^2-f_{\rm 21}^2f_{\rm 33}^2 & f_{\rm 21}^2f_{\rm 32}^2-f_{\rm 31}^2f_{\rm 22}^2 \\
f_{\rm 32}^2f_{\rm 13}^2-f_{\rm 12}^2f_{\rm 33}^2 & f_{\rm 11}^2f_{\rm 33}^2-f_{\rm 31}^2f_{\rm 13}^2 & f_{\rm 31}^2f_{\rm 12}^2-f_{\rm 11}^2f_{\rm 32}^2 \\
f_{\rm 12}^2f_{\rm 23}^2-f_{\rm 22}^2f_{\rm 13}^2 & f_{\rm 21}^2f_{\rm 13}^2-f_{\rm 11}^2f_{\rm 23}^2 & f_{\rm 11}^2f_{\rm 22}^2-f_{\rm 21}^2f_{\rm 12}^2 
 \end{pmatrix} \nonumber \\
\rm{det(F)}&=&f_{\rm 11}^2f_{\rm 22}^2f_{\rm 33}^2 + f_{\rm 21}^2f_{\rm 32}^2f_{\rm 13}^2 + f_{\rm 31}^2f_{\rm 12}^2f_{\rm 23}^2 \nonumber \\
&& - f_{\rm 11}^2f_{\rm 32}^2f_{\rm 23}^2 - f_{\rm 21}^2f_{\rm 12}^2f_{\rm 33}^2 - f_{\rm 31}^2f_{\rm 22}^2f_{\rm 13}^2
\end{eqnarray}
With the inverse of of the matrix in hand we can now use Eq.~(\ref{eq:wiiomatrix}) to write the true auto-correlations in terms of the observed auto-correlations,
\begin{equation}
\wiiT {\rm{det(F)}}=\wiio\left(\frac{\Nio}{\NiT}\right)^2(f_{\rm jj}^2f_{\rm kk}^2-f_{\rm kj}^2f_{\rm jk}^2) + \wjjo\left(\frac{\Njo}{\NiT}\right)^2(f_{\rm ki}^2f_{\rm jk}^2-f_{\rm ji}^2f_{\rm kk}^2) + \omega_{\rm kk}^{\rm o}\left(\frac{N_{\rm k}^{\rm o}}{\NiT}\right)^2(f_{\rm ji}^2f_{\rm kj}^2-f_{\rm ki}^2f_{\rm jj}^2), \label{eq:wiit}
\end{equation}
\noindent where i$\neq$j$\neq$k. The three true auto-correlation functions are found by permutations of the indicies (i,j,k)=(1,2,3), (2,3,1) and (3,1,2). Note that the equation is symmetric in the last two indicies yielding the same result for (i,j,k) and (i,k,j). Substituting these equations into Eq.~(\ref{eq:nbin_wijo}) for the observed cross-correlations we find,
\begin{eqnarray}
 \wijo\rm{det(F)}&=&\wiio\frac{\Nio}{\Njo}\left[\fii\fij(\fjj^2\fkk^2-\fkj^2\fjk^2)+\fji\fjj(\fkj^2\fik^2-\fij^2\fkk^2)+\fki\fkj(\fij^2\fjk^2-\fjj^2\fik^2)\right] \nonumber \\
&+&\wjjo\frac{\Njo}{\Nio}\left[\fii\fij(\fki^2\fjk^2-\fji^2\fkk^2)+\fji\fkk(\fii^2\fkk^2-\fki^2\fik^2)+\fki\fkj(\fji^2\fik^2-\fii^2\fjk^2)\right] \nonumber \\
&+&\omega_{\rm kk}^{\rm o}\frac{N_{\rm k}^{\rm o^2}}{\Nio\Njo}\left[\fii\fij(\fji^2\fkj^2-\fki^2\fjj^2)+\fji\fjj(\fki^2\fij^2-\fii^2\fkj^2)+\fki\fkj(\fii^2\fjj^2-\fji^2\fij^2)\right] 
\end{eqnarray}
\noindent Permuting the indicies as above yields equations for the three observed cross-correlation functions. There are three equations and six unknowns --note that $\fii$, $\fjj$ and $\fkk$ depend only on $\fij$, $\fik$, $\fji$, $\fjk$, $\fki$ and $\fkj$. By considering more than one scale we can double the number of equations making the system constrained. 

\end{document}